\documentclass[fleqn,usenatbib]{mnras}

\usepackage{newtxtext,newtxmath}
\usepackage{color}
\definecolor{referee}{rgb}{0,0,0}

\usepackage[T1]{fontenc}
\usepackage{ae,aecompl}

\usepackage{graphicx}	
\usepackage{amsmath}	
\usepackage{amssymb}	
\usepackage{lastpage}
\usepackage{hyperref}
\usepackage{supertabular,booktabs}

\title[Finding the NIR and Optical Lenses in HerBS]{A search for the lenses in the \textit{Herschel} Bright Sources (HerBS) Sample}
\author[T. J. L. C. Bakx]{Tom J. L. C. Bakx$^{1,2,3}$\thanks{E-mail: bakx@a.phys.nagoya-u.ac.jp (Nagoya University)},
Stephen Eales$^{1}$ and Aristeidis Amvrosiadis$^{1}$
\\
$^{1}$ Department of Physics and Astronomy, Cardiff University, The Parade, Cardiff, CF24 3AA, United Kingdom\\
$^{2}$ Division of Particle and Astrophysical Science, Graduate School of Science, Nagoya University, Aichi 464-8602, Japan.\\
$^{3}$ National Astronomical Observatory of Japan, 2-21-1, Osawa, Mitaka, Tokyo 181-8588, Japan.\\
}

\date{Accepted 2020 February 16. Received 2020 February 10; in original form 2019 June 27.}

\pubyear{2020}

\begin{document}
\label{firstpage}
\pagerange{\pageref{firstpage}--\pageref{lastpage}}
\maketitle

\begin{abstract}
Verifying that sub-mm galaxies (SMGs) are gravitationally lensed requires time-expensive observations with over-subscribed high-resolution observatories. 
Here, we aim to strengthen the evidence of gravitational lensing within the  \textit{Herschel} Bright Sources (HerBS) by cross-comparing their positions to optical (SDSS) and near-infrared (VIKING) surveys, in order to search for the foreground lensing galaxy candidates. 
Resolved observations of the brightest HerBS sources have already shown that most are lensed, and a galaxy evolution model predicts that $\sim$76 \% of the total HerBS sources are lensed, although with the SDSS survey we are only able to identify the likely foreground lenses for 25 \% of the sources.
With the near-infrared VIKING survey, however, we
are able to identify the likely foreground lenses for 57 \% of the sources, and we estimate that 82 \% of the HerBS sources have lenses on the VIKING images even if we cannot identify the
lens in every case. We find that the angular offsets between lens and \textit{Herschel} source are larger than that expected if the
lensing is done by individual galaxies. We also find
that the fraction of HerBS sources that are lensed falls with decreasing 500-micron flux density, which is expected from the galaxy evolution model. Finally, we apply our statistical VIKING cross-identification to the entire \textit{Herschel}-ATLAS catalogue, where we also find that the number of lensed sources falls with decreasing 500-micron flux density.
\end{abstract}



\section{Introduction}
Recent developments in far-infrared technology have allowed us to detect a population of sub-mm bright, optically faint galaxies at high redshift. These sources are forming stars at several hundreds or thousand times the typical rates of galaxies in the local Universe, and they are expected to be the progenitors of the most massive galaxies \citep{Blain2002,Casey2014}.
From 2009 until 2013, the \textit{Herschel} Space Observatory observed several large fields \citep{Pilbratt2010}, all together covering more than 1000 square degrees \citep{Eales2010,Oliver2012}, and detecting around a million new sub-mm selected sources using the Photoconductor Array Camera and Spectrometer (PACS; \citealt{Poglitsch2010}) and Spectral and Photometric Imaging Receiver (SPIRE; \citealt{Griffin2010}).

The large areas of the \textit{Herschel} surveys makes them suited for finding rare objects, such as gravitationally lensed sources. One of these large surveys is the H-ATLAS (\textit{Herschel} Astrophysical Terahertz Large Area Survey - \citealt{Eales2010}; \citealt{Valiante2016}), which covers 660 square degrees. 
Following the predictions of \cite{Negrello2007}, \cite{Negrello2010} demonstrated on the first H-ATLAS field that, once all the radio-loud AGN and nearby galaxies had been removed, all the bright ($\rm S_{500 \mu m} > 100\ mJy$) high redshift \textit{Herschel} sources are lensed. This method was used to create several samples of lensed systems. \cite{Wardlow2013} followed a similar approach on the 94 sqr. deg. HerMES (\textit{Herschel} Multi-tiered Extragalactic Survey) maps, selecting 13 sources with $\rm S_{500\mu m}$ > 100mJy. Follow-up observations of nine of the sources with the Sub-Millimetre Array (SMA), the Hubble Space Telescope (HST), Jansky Very Large Array (JVLA), Keck, and Spitzer confirmed that six of them are lensed. Recently, \cite{Negrello2016} and \cite{Nayyeri2016} used the same $\rm S_{500 \mu m} > 100\ mJy$ flux density cut-off on the full H-ATLAS and HeLMS (HerMES Large Mode Survey; 372 sqr. deg.) maps, and created samples containing 77 and 80 sources, respectively. Spectroscopic and optical follow-up observations have so far been able to confirm that 20 sources are lensed, one is a weakly-lensed proto-cluster \citep{Ivison2013}, while the remaining sources in \cite{Negrello2016} await more observations to confirm their lensing nature.

These bright \textit{Herschel} lensed sources \citep{Wardlow2013,Negrello2016,Nayyeri2016} have two practical uses. First, high-resolution observations with ALMA of these systems, aided by the magnification produced by the lensing, can provide images of the interstellar medium of the sources with a resolution as high as 30 pc \citep{ALMA2015,Dye2015,Tamura2015,Rybak2015}. Second, high-resolution images can be used to look for low-mass substructure in the lensing galaxies \citep{Vegetti2012,Hezaveh2016a,Hezaveh2016b}.

Two other potential uses of the \textit{Herschel} lensed sources are to measure cosmological parameters \citep{Eales2015} or to test the mass functions for dark-matter halos 
predicted by numerical simulations \citep{Eales2015,Aris2018}.
\cite{Eales2015}, for example, showed that observations of a sample of 100 lensed \textit{Herschel} sources would be enough to estimate $\Omega_{\Lambda}$ with a precision of 5 per cent and suggested that with observations of 1000 lensed sources it would be possible to carry out a useful investigation of the equation of state of dark energy. 
These latter two uses have so far been limited by the fact that the number of \textit{Herschel} lensed sources found by the simple technique of selecting all the high-redshift sources with 500-$\mu$m flux density $>$100 mJy is limited to $\simeq$100-200 sources \citep{Nayyeri2016,Negrello2016}, the approximate number of high-redshift sources above this flux \textcolor{referee}{density} limit.
In principle, this is not a problem because there are many more lensed sources below this flux \textcolor{referee}{density} limit, but the challenge is in finding these lensed sources among the hundreds of thousands of sources that are not lensed.

A reliable method for finding the lensed sources would be to reveal the structural signs of lensing -- arcs, Einstein rings etc. -- by making high-resolution observations with observatories such as ALMA, SMA and NOEMA. However, this is time-consuming and is not feasible for the numbers of sources found in the \textit{Herschel} surveys. \cite{GN2012} suggested an alternative approach of looking for objects in optical or near-infrared surveys that are statistically likely to be associated with the \textit{Herschel} source, but which have properties that suggest they are foreground lenses rather than the galaxies producing the sub-mm emission. The lenses and the galaxies producing the sub-mm emission are generally very different, with the latter being exceptionally luminous dusty galaxies (so-called 'sub-mm galaxies' - SMGs) at $\rm z > 2$, whereas the lenses are generally massive elliptical galaxies at $\rm z < 1$ \citep{Negrello2010,Fu2012,Wardlow2013}, which have an old stellar population and are thus very red. The red colours of the lenses makes near-infrared surveys, such as those carried out with Visible and Infrared Survey Telescope for Astronomy (VISTA), ideal for a search for the lenses.

A number of techniques have been developed for finding objects on optical surveys, such as the Sloan Digital Sky Survey, or near-infrared surveys, such as those carried out with VISTA, that are likely to be associated with a \textit{Herschel} source - whether the galaxy producing the sub-mm emission or a foreground lens - such as the probabilistic XID+ method developed for the deep \textit{Herschel} surveys \citep{Hurley2017}. The method developed by the H-ATLAS team for the wide-area surveys is the simpler one of using the magnitude and distance from a \textit{Herschel} source of a  galaxy found in an optical/near-IR catalogue to calculate the probability that the galaxy is associated with the \textit{Herschel} source (\citealt{Fleuren2012,Bourne2016,Furlanetto2018}). 
Although this method does not distinguish whether the optical/near-IR counterpart is producing the sub-mm emission itself or is a foreground lens, \cite{Bourne2014} have shown that the distribution of angular distances between the optical counterparts and the \textit{Herschel} positions is more extended for the \textit{Herschel} sources expected, on the basis of their sub-mm spectral energy distributions, to lie at higher redshifts, suggesting that a high fraction of the high-redshift \textit{Herschel} sources are lensed. \textcolor{referee}{In fact, this was already expected from the cross-correlation analysis of \cite{GN2014}. They report on the magnification bias, produced mostly by clusters of galaxies. Their subsequent study \citep{GN2017} also found the small-scale part of the cross-correlation function to be dominated by strong gravitational lensing.}

An explicit lensing search, a statistical continuation of the HALOS project of \cite{GN2012}, has also found that gravitational lens candidates do not have the same clustering properties as the background SMG distribution, but instead trace the foreground SDSS lensing distribution \citep{GN2019}. In their work, \cite{GN2019} developed a method of using the optical and sub-mm flux \textcolor{referee}{densities}, the angular distance between the optical and sub-mm positions, and the estimated redshifts of the sub-mm and optical sources to look for galaxies on the SDSS that are lensing \textit{Herschel} sources. They provide an online catalogue with the associated lensing probabilities, and find 447 sources within the area of H-ATLAS that is covered by SDSS ($\sim$340 sqr. deg.) with a $>$70\% probability of being lensed. 

In this paper, we have searched for the lenses for the \textit{Herschel} Bright Source (HerBS) sample \citep{Bakx2018} using the SDSS  and VIKING \textcolor{referee}{\citep{Edge2013}} surveys. This sample consists of the 209 brightest
($\rm S_{500 \mu m} > 80\ mJy$) sources from H-ATLAS with estimated redshifts $>$~2.
Models predict that roughly three quarters of these
sources are strongly lensed \textcolor{referee}{\citep{Negrello2007,Cai2013,Bakx2018}}. Since we expect a large fraction of
the optical/near-IR counterparts to be foreground lenses, we have adapted the standard method used to find the
optical/near-IR counterparts to H-ATLAS sources \citep{Bourne2016} to make the method more sensitive
for finding the lenses.

The paper is arranged as follows. In Section \ref{sec:sample}, we describe the sample and give an overview of the statistical method. In Section \ref{sec:sdss}, we describe the results of applying this method to the
SDSS. In Section \ref{sec:VIKING}, we apply the results of applying our method to
the near-infrared VISTA Kilo-degree Infrared Galaxy Survey (VIKING). Section \ref{sec:discussion} is a discussion, in which we compare our results against the results of other searches for lenses and discuss the possibility of applying the
method to the entire H-ATLAS survey. The conclusions are in Section \ref{sec:conclusion}.

\section{Sample and Methods}
\label{sec:sample}
\subsection{HerBS sample}
The HerBS sample consists of the brightest sources from the 660 sqr. deg. \textit{Herschel}-ATLAS survey \citep{Eales2010,Valiante2016}. HerBS galaxies are selected with SPIRE S$_{\textrm{500}\mu m}$ > 80 mJy and photometric redshifts z$_{\textrm{phot}}$~>~2. These photometric redshifts were obtained by fitting the dust spectral energy distribution that \cite{Pearson2013} 
found was a good fit to the sub-mm flux densities of high-redshift \textit{Herschel} sources to the sub-mm flux densities of each HerBS source. Local galaxies and blazars have been removed with the use of 850 $\mu m$ SCUBA-2 observations \citep{Bakx2018}. The sample of \cite{Negrello2016},
a sample of potential lensed sources with S$_{\rm{500}\mu m}$ $>$ 100 mJy,
has 63 sources in common with HerBS. 

\textcolor{referee}{\cite{Cai2013} presented a detailed model of galaxy evolution in the infrared and sub-mm wavebands, including the effect of gravitational lensing.  The galaxy evolution model by \cite{Cai2013} builds upon the initial physical models in \cite{Granato2004}, and the initial implementations of these models to observational data in \cite{Lapi2006,Lapi2011}.}
The model of \cite{Cai2013} predicts that $\simeq$76\% of the HerBS sample are lensed Ultra Luminous InfraRed Galaxies (ULIRGs; L$_{\rm{FIR}}$ = 10$^{12}$ - 10$^{13}$ L$_{\odot}$), with the other sources being unlensed Hyper-Luminous InfraRed Galaxies (HyLIRGS; L$_{\rm{FIR}}$ $>$ 10$^{13}$ L$_{\odot}$). The model predicts that virtually all the sources with $\rm S_{500\mu m}>100\ mJy$ are lensed \textcolor{referee}{(as already predicted by \citealt{Negrello2007})}, with the lensing fraction falling rapidly below this flux density. 
The sources fall in five fields: two fields near to the North and South Galactic Poles (NGP and SGP, respectively), and three equatorial fields that are the same as the 9-hour, 12-hour and 15-hour fields observed as part of the GAMA redshift survey (henceforth GAMA09, GAMA12 and GAMA15; \citealt{Driver2011}).

\subsection{The identification method}
\label{sec:likelihood}
In our standard method we use a Bayesian technique to estimate the
probability of a potential counterpart to the \textit{Herschel} source on an
optical/near-IR image is actually associated with the sources. The method
uses the magnitude ($m$) and angular distance ($r$) 
to the \textit{Herschel} source of the
potential counterpart and is described in detail in
\cite{Sutherland1992}, with the H-ATLAS version of the technique described
in \cite{Bourne2016}.

For a possible counterpart to a (\textit{Herschel}) source the first step is to calculate the ratio of two likelihoods:
\begin{equation}
L = \frac{q(m) f(r)}{n(m)}.
\label{eq:likelihood}
\end{equation}
The numerator is the probability of obtaining a counterpart with the
the measured values of $m$ and $r$ on the assumption that the potential
counterpart is actually associated with the \textit{Herschel} source, whether
as the galaxy producing the submillimetre emission or as a lens. The denominator
is the probability of obtaining the measured values of $m$ and $r$ on the assumption that the object is completely unrelated to the \textit{Herschel} source.
$q(m)$ represents the probability distribution for the magnitudes of genuine counterparts, $n(m)$ represents the background surface density distribution of unrelated objects (in units of arcsec$^{-2}$) and $f(r)$ represents the distribution of offsets between sub-mm and near-IR positions produced by both positional errors between both catalogues and gravitational lensing offsets (in units of arcsec$^{-2}$). 

All components are probability distributions, and so $q(m)$ should, integrated over all magnitudes, equal the probability that a \textit{Herschel} source is detected in the optical/near-IR survey, which we will call $Q_{0}$. Hence,
\begin{equation}
Q_0 = \int_{-\infty}^{\infty} q(m)dm.
\end{equation}
Similarly, the integral of the positional offset surface density distribution, $f(r)$, over all available area should equal 1, and so
\begin{equation}
1 \equiv \int_A f(r) dA = \int_0^{2\pi} \int_0^{\infty} f(r) r dr d\phi = \int_0^{\infty} 2\pi r f(r) dr.
\end{equation}
$n(m)$ is the surface-density of galaxies that are unrelated to the \textit{Herschel} sources as a function of magnitude and can be estimated from
\begin{equation}
n(m) = \frac{n_{\textrm{back}}(m)}{\textrm{Area}} = \frac{n_{\textrm{back}}(m)}{\pi r^2 N_{\textrm{random}}}.
\label{eq:nback}
\end{equation}
in which $n_{\textrm{back}}(m)$ 
is the histogram of magnitudes for galaxies within 10 arcsec of a set of random positions on the optical/near-IR images, Area
is the total search area and $N_{\textrm{random}}$ is the number of
random positions.

We estimate the actual probability (the 'reliability') that
a potential counterpart is associated with a (\textit{Herschel}) source
by the weighted combination of the likelihood ratios of all potential counterparts near to that (\textit{Herschel}) source:
\begin{equation}
R_j = \frac{L_j}{\sum_i L_i + (1-Q_0)}.
\label{eq:Reliability}
\end{equation}
The reliability $R_{j}$ of each potential match, $j$, is computed as the ratio of its likelihood ratio ($L_j$) to the sum of the likelihood ratios of all potential matches within 10 arcseconds. An extra term in the denominator, $(1 - Q_0)$, accounts for the possibility that the optical/near-IR counterpart
to the \textit{Herschel} source is not visible on the images. 
In previous work on finding the counterparts to H-ATLAS sources,
\cite{Bourne2016} found $Q_0$ = 0.519 for the SDSS r-band images and \cite{Fleuren2012} found $Q_0$ = 0.7342 $\pm$ 0.0257 in a pilot analysis with
VISTA K-band images. 
Throughout this paper, we count an optical/near-IR counterpart with R $>$ 0.8
as a probable counterpart, although we note that this only represents an 80\% probability.

\section{SDSS}
\label{sec:sdss}
\subsection{Previous Results}
\cite{Bourne2016} and \cite{Furlanetto2018} 
have looked for optical counterparts to the H-ATLAS sources
on the SDSS r-band images, the former for the GAMA fields and
the latter for the NGP (the SGP field
is not covered by the SDSS). 121 HerBS sources fall in these fields, 72 in GAMA
and 49 in NGP. Only 31 of the sources (25\%) have a counterpart with R$>$0.8
in either catalogue, much lower than 76 \% predicted by the galaxy evolution model of \cite{Cai2013}. On the assumption the model is correct, only 34 \% of the expected lenses are found in these catalogues. If the model is correct, one
possible explanation is that we are missing the lenses because they are
too faint to be detected on the SDSS r-band images, which is plausible because
the lenses in some of the \textit{Herschel} lensed systems are at $z>1$
\citep{Cox2011,GN2012,Bussmann2013}. A second possibility, which we consider in the nex section, is that the lenses are on the SDSS images but too far from
the \textit{Herschel} sources to have $R>0.8$ and thus be classified by
us as counterparts.

\subsection{Toy Model: Ad hoc inclusion of gravitational lensing}
\label{sec:toymodel}
A possible cause of the low SDSS detection fraction 
is if the lenses are too far from the \textit{Herschel} sources for 
our method to find a reliable counterpart (R~$>$~0.8). 
This might also explain the large offsets between \textit{Herschel} sources
and lenses found for high-redshift \textit{Herschel} sources
by \cite{Bourne2014}. 

We have tested this hypothesis by including in $f(r)$ (Equation 1) an additional
term produced by gravitational lensing on top of the usual term that
arises from the uncertain positions of the \textit{Herschel} sources.
We used a Monte Carlo simulation to generate a distribution of $f(r)$ that 
includes the effects of astrometric errors, using the Gaussian distribution from \cite{Bourne2016}, and the empirical image separations found for a sample of lensed systems by \cite{Aris2018}. 

To produce the astrometric errors, we used a Gaussian probability
distribution with $\sigma_{\rm pos}$ = 1 arcsec \citep{Bourne2016}, which is true for sources detected at more than $\sim$10$\sigma$ at 250 $\mu$m, and thus holds true for almost all HerBS sources. We derived a probability distribution for the
lensing offsets from the histogram of Einstein radii found by
\cite{Aris2018} from ALMA observations of 15 bright \textit{Herschel} sources.
We use these probability distributions to generate 10\,000\,000 combined angular distances, calculating a set of offsets in both the x- and y-direction that include both astrometric errors and lensing, $\theta_{\textrm{tot}}$. We then use this distribution of combined angular distances to generate a probability distribution by normalizing the combined distribution to 1, integrated over all area, as it is a surface probability distribution:
\begin{equation} 
1 = \int f\left(\sqrt{x^2 + y^2}\right)dA = \int f(r')dA = \int_0^{\infty} f(r') 2\pi r' dr'.
\label{eq:normalizationOfDistribution}
\end{equation}

\begin{figure}
  \includegraphics[width=0.5\textwidth]{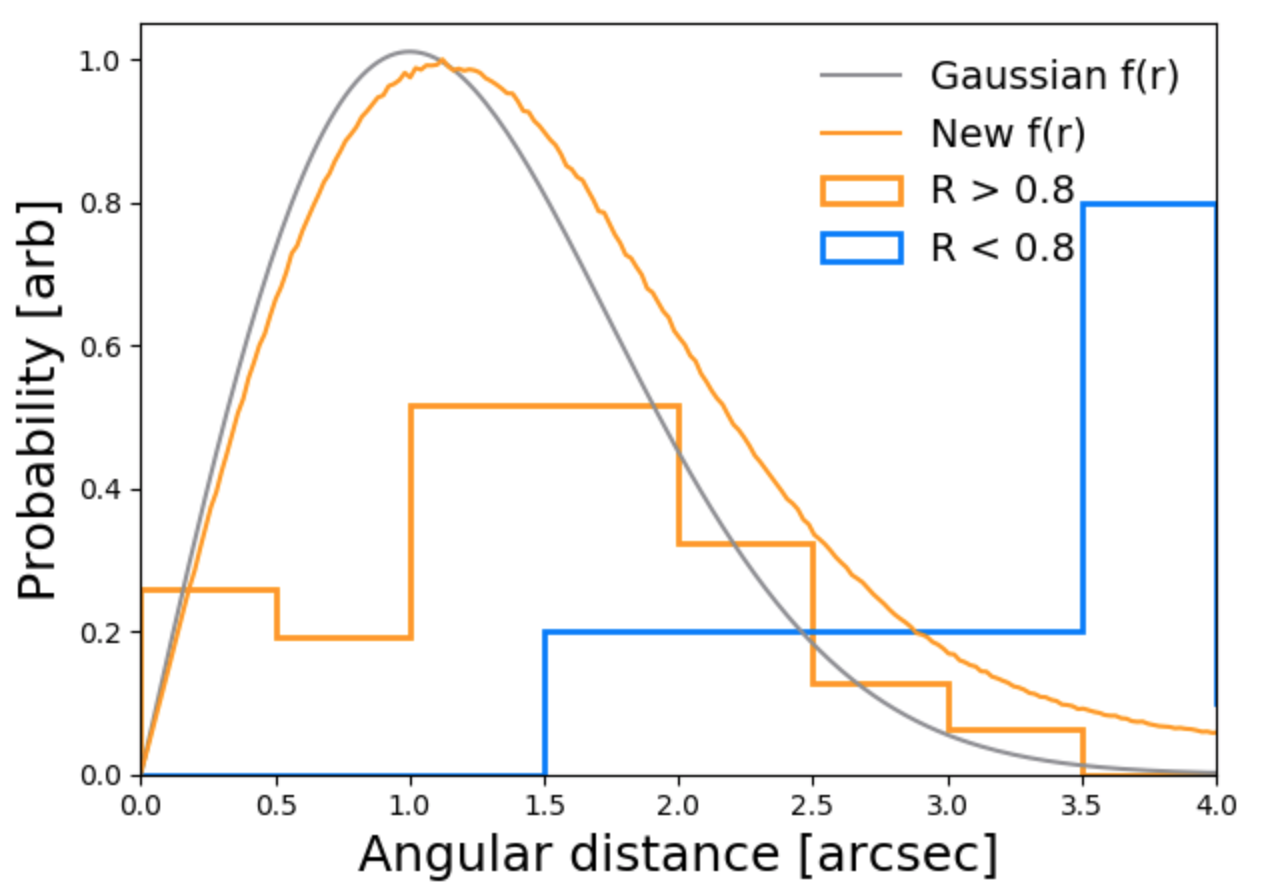}
  \caption{The old (\textit{grey}) and new (\textit{orange line}; toy-model) radial distribution functions are compared against the radial distribution of R < 0.8 (\textit{blue histogram}) and R > 0.8 SDSS counterparts (\textit{orange histogram}). The old and new radial distribution functions vary slightly, however the new angular distribution extends further due to gravitational lensing. The histogram of sources with R > 0.8 agrees poorly with the old and new angular distributions at low angular distances, however the distribution seems to agree with the tail end of our new distribution. The y-scale of the histograms are adjusted to better match the radial distribution functions.}
  \label{fig:MCs}
\end{figure}
In Figure \ref{fig:MCs} we compare our new angular distribution (\textit{orange line}) to the distribution of angular offset for counterparts with R < 0.8 (\textit{blue histogram}) and for counterparts with R > 0.8 SDSS counterparts (\textit{orange histogram}). To do so, we need to convert the angular distribution function from a surface probability ($f(r)$ in units of arcsec$^{-2}$) to a radial distribution ($f^\dagger(r)$ in units of arcsec$^{-1}$), using 
\begin{equation}
f^\dagger(r) \equiv (dA / dr) f(r) = 2 \pi r f(r).
\end{equation}

A comparison of the predicted distribution with the distribution 
for the counterparts with R > 0.8  gives poor agreement at low angular distances ($\theta$ < 1 arcsec) but better agreement at higher angular distances.

\subsection{The effect of gravitational lensing on the SDSS likelihood analysis}
\label{sec:GravLensingOnSDSS}
We estimate the missed number of gravitational lenses in the SDSS likelihood analysis with two separate methods that use the actual images of lensed \textit{Herschel} sources from \cite{Aris2018}. We used the toy model
in the previous section to estimate the increase in the distance between counterpart and \textit{Herschel} source produced by the lensing in addition to the effect of astrometric errors (Figure 1). This will allow us to replace the $f(r)$ in equation \ref{eq:likelihood}, in order to understand the effects gravitational lensing has on our method of finding the SDSS counterparts to the \textit{Herschel} sources.

We use two methods for calculating the total number of missed counterparts, each described in detail below. We define the missed counterparts as the sources with SDSS galaxies too far from the \textit{Herschel} position to have a reliability greater than 0.8. In the first method, we use the new $f(r)$ to calculate the number of missed counterparts analytically. In the second method, 
we repeat the entire method used by \cite{Bourne2016} and \cite{Furlanetto2018} using
our new version of $f(r)$ and recalculating the reliability for each potential counterpart.

\subsubsection{First method: Statistical approach}
This statistical approach uses all the counterparts that are reliabily detected, with an R $>$ 0.8. From the likelihood value of this counterpart, $L$, we can calculate the maximum radius until which this source would have an R equal to 0.8 with the original radial probability distribution. We can then use the radial distribution that includes gravitational lensing to calculate the probability that this source would have been located outside of this region, and would thus have an R $<$ 0.8.

We use an example to illustrate why \textcolor{referee}{this} approach works: if an actual counterpart source, with R $>$ 0.8, has only a 20\% chance to lie within the detectable area due to an additional angular offset caused by gravitational lensing, this indicates that for each counterpart we identify, we will on average have missed four (= (1/0.2) - 1).

Mathematically, we calculate the total number of missed sources by summing the inverse of the probability it was detected,
\begin{equation}
     N_{missed} = \sum^i \left[\frac{1}{\int_{0}^{r_{lim,i}} p_{tot}(r) dr} - 1 \right].
     \label{eq:missed}
\end{equation}
Here we sum over $i$ for every source with R > 0.8, $r_{lim,i}$ refers to the maximum angle at which R = 0.8, and $p_{tot}(r)$~$\delta r$ is the probability that the counterpart will lie at
a distance $r$ within a small shell $\delta r$ from \textit{Herschel} position, a probability that contains both the effect of astrometric errors and lensing.

We calculate the total probability, $p_{tot}(\theta)$, by normalizing the new angular separation distribution from the Monte-Carlo method (shown in Figure \ref{fig:MCs}) to unity. We calculate maximum angle, $r_{lim}$, analytically by rewriting equations \ref{eq:likelihood} and \ref{eq:Reliability},

\begin{eqnarray*}
        R_j &=& \frac{L_j}{\sum^i L_i + (1 - Q_0)}.
        \end{eqnarray*}
We divide up the sum over all likelihoods, $\sum^i L_i$, into two components, namely the main component, $L_j$, and the other contributing likelihoods, $S$.
        \begin{eqnarray*}
        \sum^i L_i &=& S + L_j.
        \end{eqnarray*}
We introduce this convention into equation \ref{eq:Reliability},
        \begin{eqnarray*}
        R_j &=& \frac{L_j}{ S + L_j + (1 - Q_0)}.
        \end{eqnarray*}
We now imagine a source at the reliability inclusion limit, $R_{lim}$, which we set to 0.8,
        \begin{eqnarray*}
        R_{lim} &\equiv& 0.8.
        \end{eqnarray*}
Using this, we calculate the likelihood of this imaginary source, at which the reliability equals 0.8, $L_{lim}$,
        \begin{eqnarray*}
        0.8 &=& \frac{L_{lim}}{S + L_{lim} + 1 - Q_0}, \\
        0.2 L_{lim} &=& 0.8(S + 1 - Q_0), \\
        L_{lim} &=& 4(S + 1 - Q_0).
        \end{eqnarray*}
The likelihood can then be described as a product of the likelihood value if the source were located at an angular offset $r=0$, $L_{r=0}$, and a separate component, accounting for the angular offset distribution at $r=r_{lim}$, the distance at which the reliability is $R_{lim}$ (e.g. 0.8),
        \begin{eqnarray*}
        L_{lim} &=& L_{r=0}e^{-r_{lim}^2/2\sigma_{pos}^2}.
        \end{eqnarray*}
        Here, $\sigma_{pos}$ is the standard deviation of the Gaussian distribution that represents the astrometric errors.
Finally, we can compare this imaginary source with a reliability of $R_{lim}$ to a real source, with a measured likelihood, $L_{meas}$, and angular offset, $r_{meas}$,
        \begin{eqnarray*}
        L_{meas} &=& L_{r=0}e^{-r_{meas}^2/2\sigma_{pos}^2},\\
        L_{lim} &=& L_{meas}e^{(r_{meas}^2 -r_{lim}^2)/2\sigma_{pos}^2},\\
        e^{(r_{meas}^2 -r_{lim}^2)/2\sigma_{pos}^2} &=& 4 \frac{(S + 1 - Q_0)}{L_{meas}},    \\  
        r_{lim} &=& \sqrt{-2 \sigma_{pos}^2 \ln\left({ 4\frac{(S + 1 - Q_0)}{L_{meas}}}\right) + r_{meas}^2}.
\end{eqnarray*}

\begin{figure*}
\begin{minipage}{0.47\linewidth}
  \includegraphics[width=0.9\textwidth]{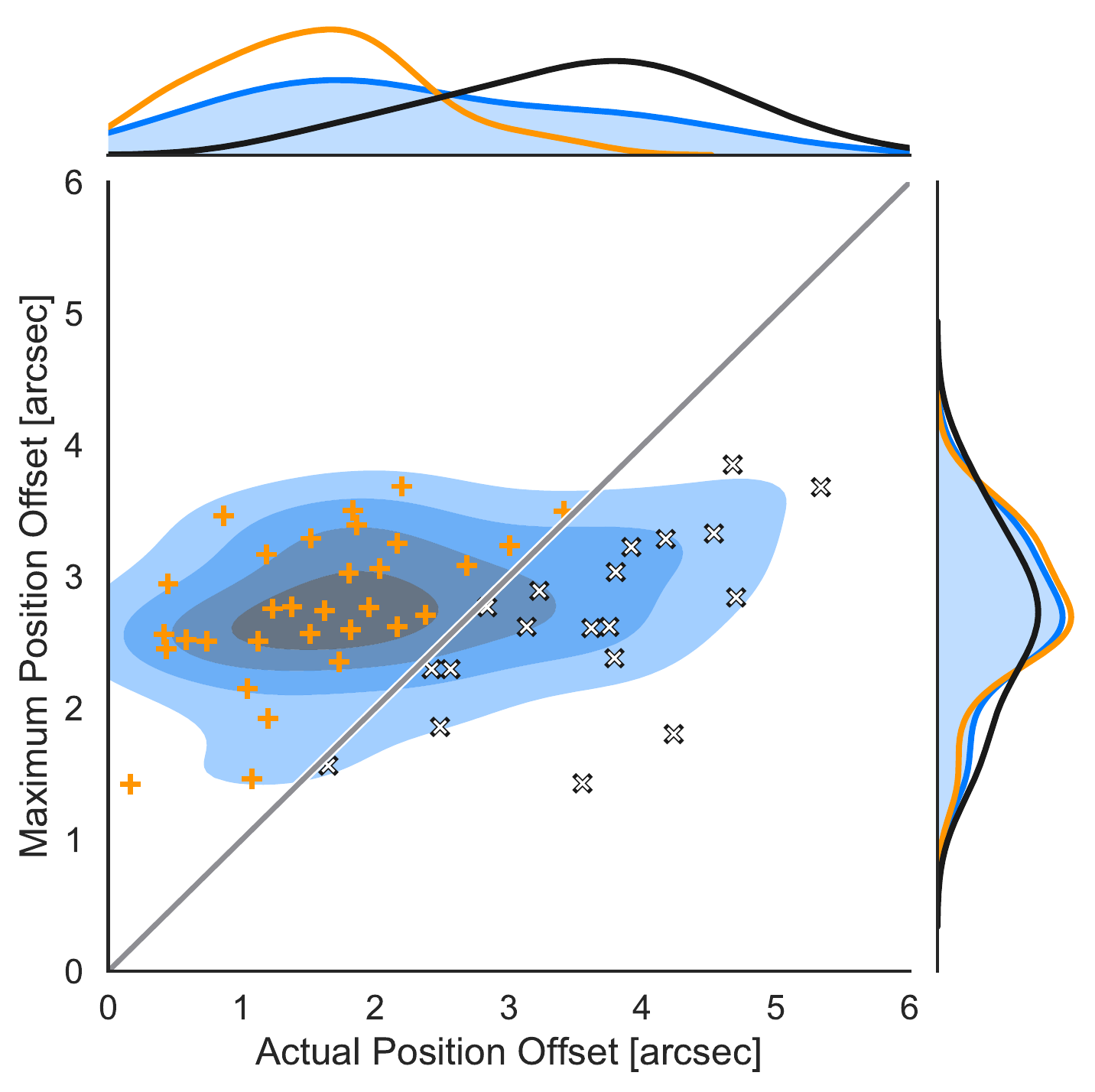}
  \caption{The actual distance between the VIKING galaxy and the \textit{Herschel} position (x-axis) plotted against the maximum distance at which the galaxy could have been revealed by our method as a counterpart to the \textit{Herschel} source (y axis). At this maximum angle, the reliability is equal to 0.8, which is the threshold for a reliable and unreliable detection. We see that R > 0.8 counterparts (\textit{orange plusses}) lie on the left-hand side of the y = x line, whilst R < 0.8 counterparts (\textit{black and white crosses}) lie on the right-hand side. This is as expected, as on this line, the reliability of the counterparts would be 0.8. We note many R < 0.8 sources close to the y = x line (\textit{grey line}). The \textit{blue} background indicates the bivariate kernel density function of all SDSS galaxies, and the top and right graphs indicate the distribution collapsed in either the maximum or actual position offset, respectively. The \textit{orange}, \textit{black} and \textit{blue} lines in the side panels refer to R > 0.8, R < 0.8 and all counterparts, respectively.}
  \label{fig:ActVsMax}
\end{minipage}
\begin{minipage}{0.05\linewidth}
  $ $
\end{minipage}
\begin{minipage}{0.47\linewidth}
  \includegraphics[width=0.9\textwidth]{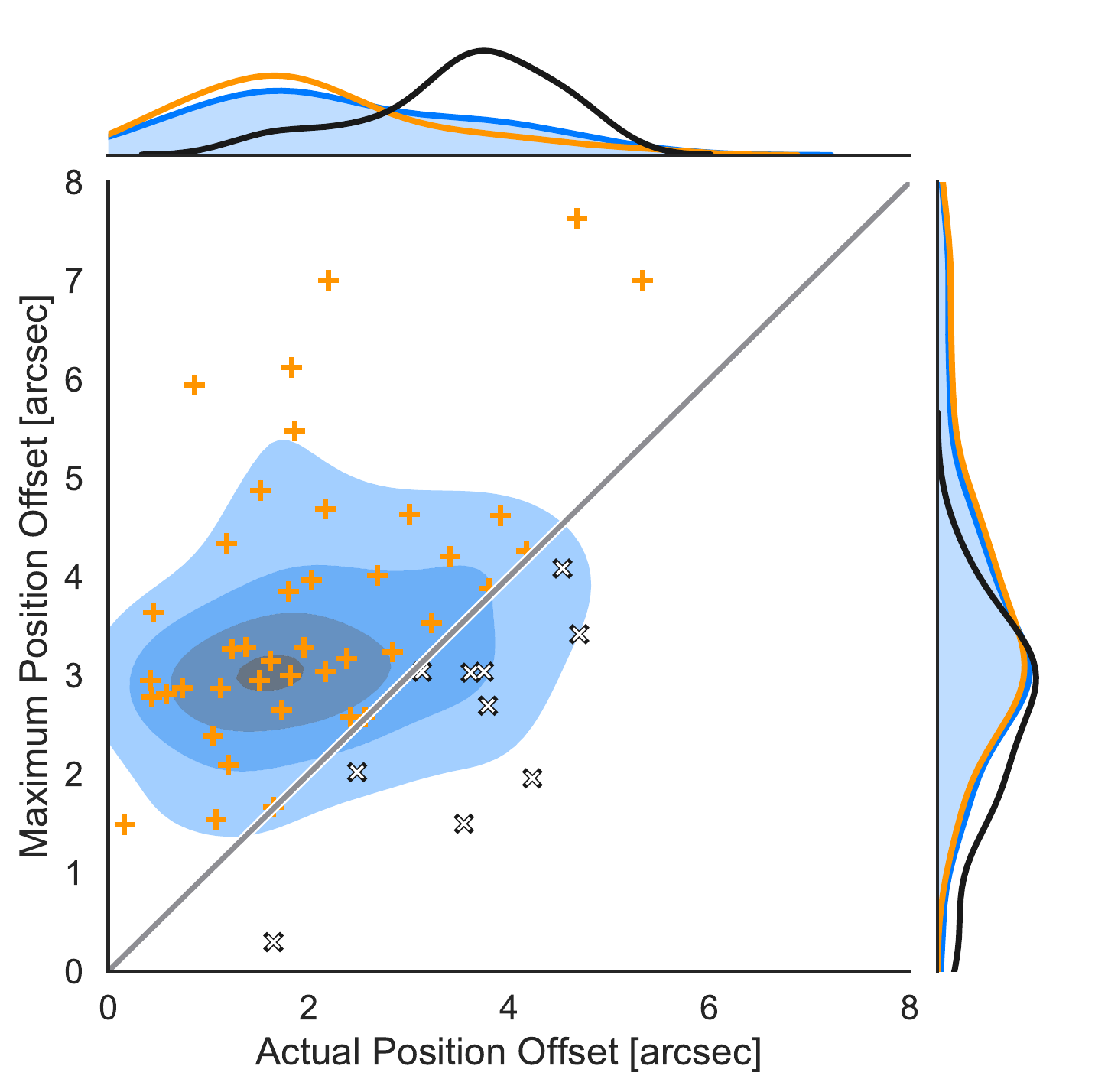}
  \caption{The same as Figure \ref{fig:ActVsMax} except that the reliabilities (R) of the counterparts are now those calculated using the method described in Section \ref{sec:secondmethod}. We note that several cross-correlations are at large distances, beyond even 6 arcseconds.}
  \label{fig:NewMax}
\end{minipage}
\end{figure*}
\begin{figure}
    \includegraphics[width=0.5\textwidth]{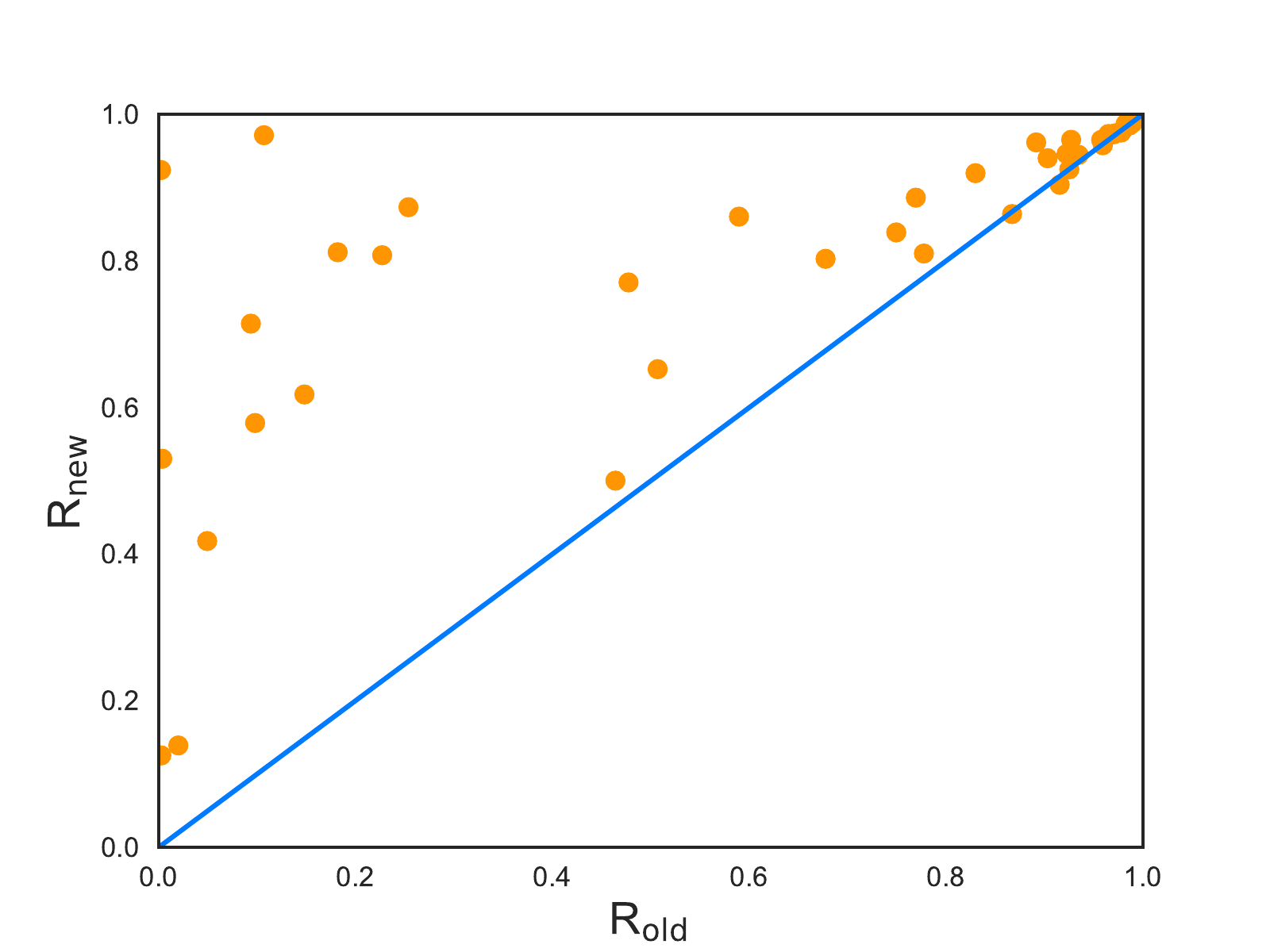}
    \caption{Reliabilities (R) calculated using our new version of $f(r)$, which includes the effect of gravitational lensing, plotted against the reliabilities given in the catalogue of \citet{Bourne2016}. Even low original reliabilities can result in reliabilities greater than 0.8 with our new analysis. The \textit{blue line} is the y = x line.}
    \label{fig:Rold}
\end{figure}
We plot $r_{lim}$ against $r_{meas}$ in Figure \ref{fig:ActVsMax}. The figure also shows the y = x line, where the measured angular separation and the maximum angular separation are equal. A source on this line would thus have a relability of 0.8. As can be expected, the R > 0.8 counterparts lie on the left-hand-side of the y = x line, while R < 0.8 counterparts lie on the right-hand-side of this line. We note a surplus of R < 0.8 counterparts close to the line. These will not be used in the calculations.

With equation \ref{eq:missed} we calculate that 1.5 lenses were missed by being too far from the \textit{Herschel} position or by having too faint a magnitude.

\subsubsection{Second method: Recalculating the reliability}
\label{sec:secondmethod}
In our other method, we calculate the number of missed sources by recalculating the reliability for all sources with documented reliabilities in \cite{Bourne2016}, but using the form of $f(r)$ corrected for the effect of lensing using the simple model described in the previous section. In the catalogues from \cite{Bourne2016}, several sources have reliability values of the order 10$^{-12}$. We discarded these sources, since such small values are easily influenced by rounding errors, and might produce spurious results. We change $f(r)$ in equation \ref{eq:likelihood} but we keep everything else (i.e. $q(m)/n(m)$) the same.

We use the original $f(r)$, a Gaussian distribution, to calculate the likelihood a galaxy would have had if it was at the position of the \textit{Herschel} source, $L_{r=0}$. This is done by simply dividing by the value of $f(r)$ for the measured angular separation. This gives us the likelihood value in case there were no angular separation, $L_{r=0}$. After this we recalculate the likelihood using the actual position of the galaxy and the corrected version of $f(r)$ that we obtained in the previous section. Then we calculate the reliability using equation \ref{eq:Reliability}.

Our re-analysis shows that we now find 41 sources with R $>$ 0.8, suggesting our initial estimate missed 10 SDSS counterparts due larger-than-expected angular separations. Figure \ref{fig:Rold} shows the new reliabilities plotted against the old reliabilities. It shows that almost all reliabilities increase, and that even very low reliabilities can be increased to above R$_{\textrm{new}}$ > 0.8.

Even with 10 missed lenses, the total number of lenses found on the SDSS is well below the number expected from the model ($\sim$90 sources across the 121 sources with SDSS coverage; \citealt{Bakx2018}). One obvious possibility, which we test in the next section, is that many of the lenses are too faint to be detected on the SDSS.

The disagreement between the two methods is large, with one method predicting 32.5 and the other 41 lensing candidates. Since the first method only uses the initial candidates, it does not account for many sources with a reliability close but less than 0.8, which result in likely candidates in the second method.
This can be seen in Figures \ref{fig:ActVsMax} and \ref{fig:NewMax}, which show the angular positional offsets between the sources, and the maximum distance at which the galaxy could have been revealed by our method as a counterpart to the \textit{Herschel} source. 
The disagreement means that the toy-model angular distribution function, together with the initial candidates, fail to represent the lensing behaviour of our sample. We will discuss this in more detail in Section \ref{sec:discAng}.
Figure \ref{fig:NewMax} is the same as Figure \ref{fig:ActVsMax} except we have now used our recalculated reliabilities for the lensing-adjusted angular separation distribution.
We note a significant increase in the maximum position offset, allowing us to cross-identify sources beyond 6 arcseconds, whereas the maximum position offset in the original method was only up to 4 arcseconds. \textcolor{referee}{These large angular separations are expected in the case of gravitational lensing, as they were predicted from the cross-correlation analysis in \cite{GN2014} and \cite{GN2017}, and noticed in \textit{Herschel} measurements of red sources by analysis in \cite{Bourne2014}}.

\section{VIKING source extraction and cross-identification}
\label{sec:VIKING}
The foreground lensing sources are expected to be red-and-dead galaxies, with evolved stellar populations, emitting mostly in the near-infrared. As the SDSS might not be detecting all the foreground lenses, we use the deeper, near-infrared VIKING survey to look for the lenses. 

\subsection{VIKING survey and source extraction}
The VISTA telescope observed 1500 square degrees for the VISTA Kilo-degree Infrared Galaxy (VIKING) survey in zYJHK$_{\rm{S}}$ to sub-arcsecond resolution, to an AB-magnitude 5$\sigma$ point source depth of 23.1, 22.3, 22.1, 21.5 and 21.2, respectively \citep{Edge2013}. The overlap with the equatorial GAMA fields and the South Galactic Pole (SGP) fields makes this survey ideal for looking for counterparts to the H-ATLAS sources. 
Currently, no VIKING catalogues exist that use the full available data for SGP and GAMA fields. \textcolor{referee}{An initial VIKING cross-correlation study was done on part of the GAMA09 field by \cite{Fleuren2012}. They studied the field covered by the Science Demonstration Phase of H-ATLAS \citep{Eales2010}, which has both been observed in the VIKING and SDSS surveys. They do not, however, provide a catalogue of the entire VIKING survey.}
The catalogue in \cite{Wright2018} only covers 28 out of the 98 HerBS sources with VIKING coverage in the current SGP and GAMA data releases. The catalogue of \cite{Wright2018} only uses the data from the third data release of the VIKING sister-survey Kilo-Degree Survey (KiDS; \citealt{deJong2017}), and it stringently removes regions close to bright stars, resulting in a Swiss-cheese-like map.
Therefore, we will identify the VIKING galaxies using SExtractor. The VIKING data used in this section comes in the form of SWarped images \citep{Bertin2010}, both an image and a weight file. The exact data reduction procedure for these images is described in \cite{Driver2016}. 

These SWarped images only cover a part of the H-ATLAS area surveyed by VIKING,  thus requiring a future analysis for a more complete picture when all data becomes available.
Two HerBS sources in GAMA09 and 60 sources in the
SGP do not fall in the region covered by these images.
In total, 98 HerBS sources are covered by the current SWarped images, and our analysis is restricted to these 98 sources. 

\subsubsection{Source-Extractor set-up}
We are interested accurately finding the lensing galaxies, which will be either unresolved or only slightly resolved in the VIKING survey and so we optimise the source extraction to create a robust catalogue of VIKING galaxies with the most accurate flux \textcolor{referee}{density} estimates for point sources. This means we are not looking to 'push' our source extraction to low flux \textcolor{referee}{densities}, with the inherent risk of including noisy features as sources. We also do not aim to extract the most accurate integrated flux \textcolor{referee}{densities} for extended sources. 

From previous studies, we know that foreground, lensing sources lie at redshifts between z = 0.1 and 1.5 \citep{Cox2011,Fu2012,Bothwell2013}. In order to probe the highest redshifts, we select sources at the longest wavelength band of the VIKING fields, K$_{\rm{S}}$ at $\sim$ 2.1 $\mu$m. We use the \textit{dual-mode} extraction, where sources are detected on the K$_{\rm{S}}$ image, and the photometry is measured on the other images (z, Y, J, H) using the source parameters provided from the K$_{\rm{S}}$ image, using the standard AUTO flux \textcolor{referee}{density} extraction.
We chose the Sextractor parameters by applying Sextractor to five test fields, adjusting the parameters until there were  no obviously real sources on the source-subtracted images.
The size of each field was 1000 by 1000 arcseconds, large enough to include a diverse selection of sources and survey properties (e.g. point sources, extended features, bright stars and overlapping mapping regions). Further details of our application of SExtractor and the final images of the five test fields can be found in the \textcolor{referee}{online supplementary material}.

\begin{figure}
\includegraphics[width=\linewidth]{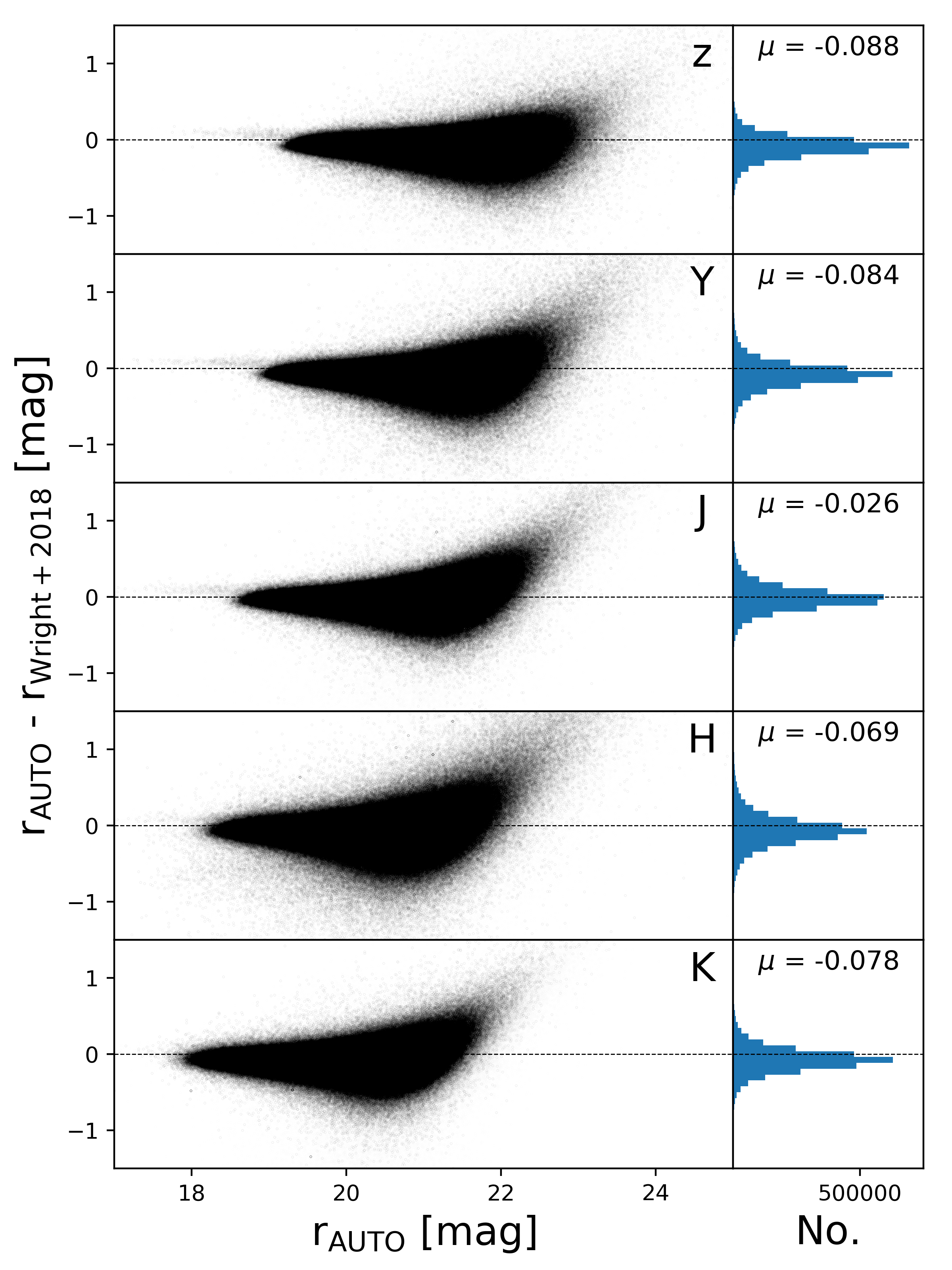}
\caption{We compare the extracted magnitude between our catalogue and the catalogue of \citet{Wright2018}. We find, on average, slightly lower magnitudes than \citet{Wright2018}, although at higher magnitudes, we seem to find higher magnitudes than \citet{Wright2018}. $\mu$ refers to the average difference in magnitude between our catalogue and \citet{Wright2018}. The small dispersion suggests our extraction is of sufficient quality for our analysis. }
\label{fig:WrightSExtractor}
\end{figure}

We compare the quality of our extraction to the catalogue of \cite{Wright2018} in Figure \ref{fig:WrightSExtractor}. Here, we compare the magnitude of each extracted source, to the magnitude found by \cite{Wright2018}, for all five bands. On average, we find slightly lower magnitudes than \cite{Wright2018}, although for higher magnitudes, we seem to find higher magnitudes than \cite{Wright2018}. The small dispersion suggests our extraction is of sufficient quality for our analysis. 

\subsubsection{Star/Galaxy separation}
\begin{figure}
\includegraphics[width=\linewidth]{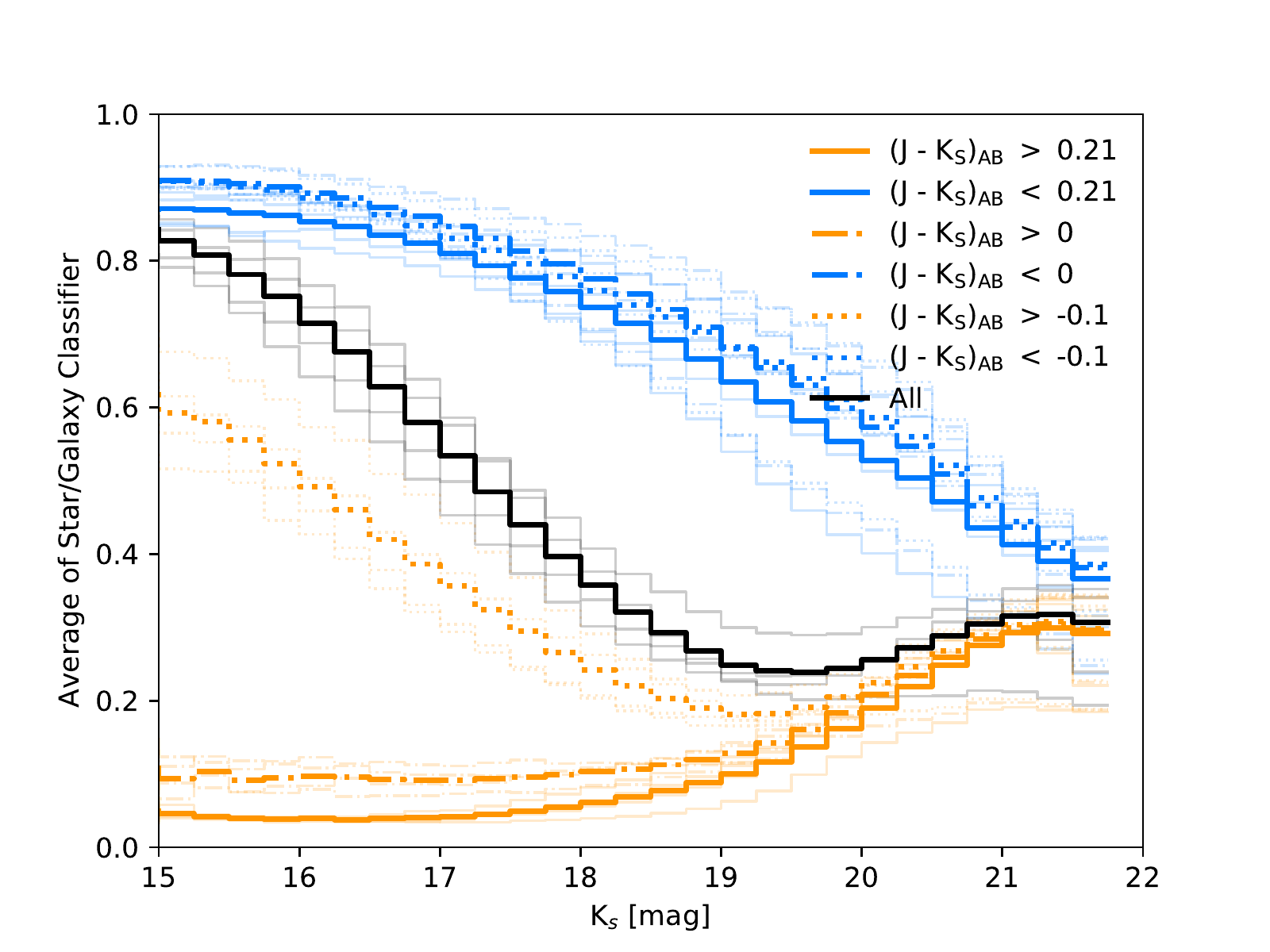}
\caption{In each flux \textcolor{referee}{density} bin, we average the star/galaxy estimator from SExtractor for three different J - K colour cuts from \citet{Fleuren2012}: J - K$_{\rm{S}}$ $>$ -0.1, $>$ 0 and $>$ 0.21. Low values of the star/galaxy estimator suggest it is more likely to be a galaxy, while values close to one suggest it is more likely to be a star. All orange lines refer to the greater-than colour cut ($>$), while all blue lines correspond to the less-than ($<$) colour cut. The black line looks at all sources. The thin lines correspond to the analysis on the individual fields (GAMA-9, -12, -15 and SGP). While the J - K $>$ 0.21 shows the most reliable cut for a galaxy selection cut, we will be removing a subset of galaxies. Therefore, we choose the J - K$_{\rm{S}}$ $>$ 0 selection. All colour cuts converge to a similar value for the star/galaxy estimator at high magnitudes, potentially because the star/galaxy classifier fails for faint sources.}
\label{fig:GalaxyOrStar}
\end{figure}
A significant number of sources in the VIKING fields are stars. These interlopers would reduce the power of our method for finding counterparts to the \textit{Herschel} sources. 
\cite{Fleuren2012} removed these stellar interlopers by means of a colour cut using SDSS and VIKING flux \textcolor{referee}{densities}. They studied the GAMA09 field as a part of a precursor study, which also had coverage from the SDSS. Not all our fields have coverage from the SDSS, so we have tried to remove stars using a single VIKING colour. Moreover, in Section \ref{sec:sdss} we found that not all lenses might be visible in the SDSS.

We examine a VIKING flux \textcolor{referee}{density}-cut for the galaxy selection based on the method of \cite{Fleuren2012}. They employed a colour-cut ranging from J - K$_{\rm{S}}$ > -0.34 to 0.21 as a function of increasing g~-~i SDSS colour. 
In Figure~\ref{fig:GalaxyOrStar}, we show the average star/galaxy estimator from Sextractor for three different cuts in J~-~K colours. This star/galaxy criterion value is supplied by SExtractor for each source, and is based on a neural network-derived value for the probability of an object being a star or galaxy. We choose three colour-cuts: J - K$_{\rm{S}}$ $>$ -$0$.1, $0$ and $0$.21.
The figure shows a similar separation ability for both the $0$ and 0.21 colour cuts, however the -0.1 colour cut clearly leads to the inclusion of a significant number of stars. From \cite{Bertin1996}, we know that the neural network is more accurate for brighter sources, and it becomes less accurate at higher magnitudes. This seems to explain why all colour cuts lead to a similar value for star/galaxy estimator for faint sources.

We choose to use the J - K$_{\rm{S}}$ $>$ $0$ colour cut, as it includes as many galaxies as possible, without an excessive inclusion of stars, however we still keep in mind that we might be excluding potential galaxies throughout the analysis.

\subsection{Implementation of the Likelihood method}
We now use this catalogue of galaxies to carry out a similar search for counterparts to that described by \cite{Bourne2016} and Section \ref{sec:sample} but with one big difference. Because we suspect from a model \citep{Cai2013}, statistics \citep{GN2017} and resolved observations \citep{Bothwell2013,Negrello2016,Aris2018} that many of the sources are lensed, we cannot assume that the offsets between the \textit{Herschel} and near-IR positions have the Gaussian distribution estimated from the astrometric errors. Instead, we calculate the distribution of offsets from the data themselves.

\subsubsection{Estimation of $Q_0$}
\begin{figure*}
\centering
\includegraphics[width=\linewidth]{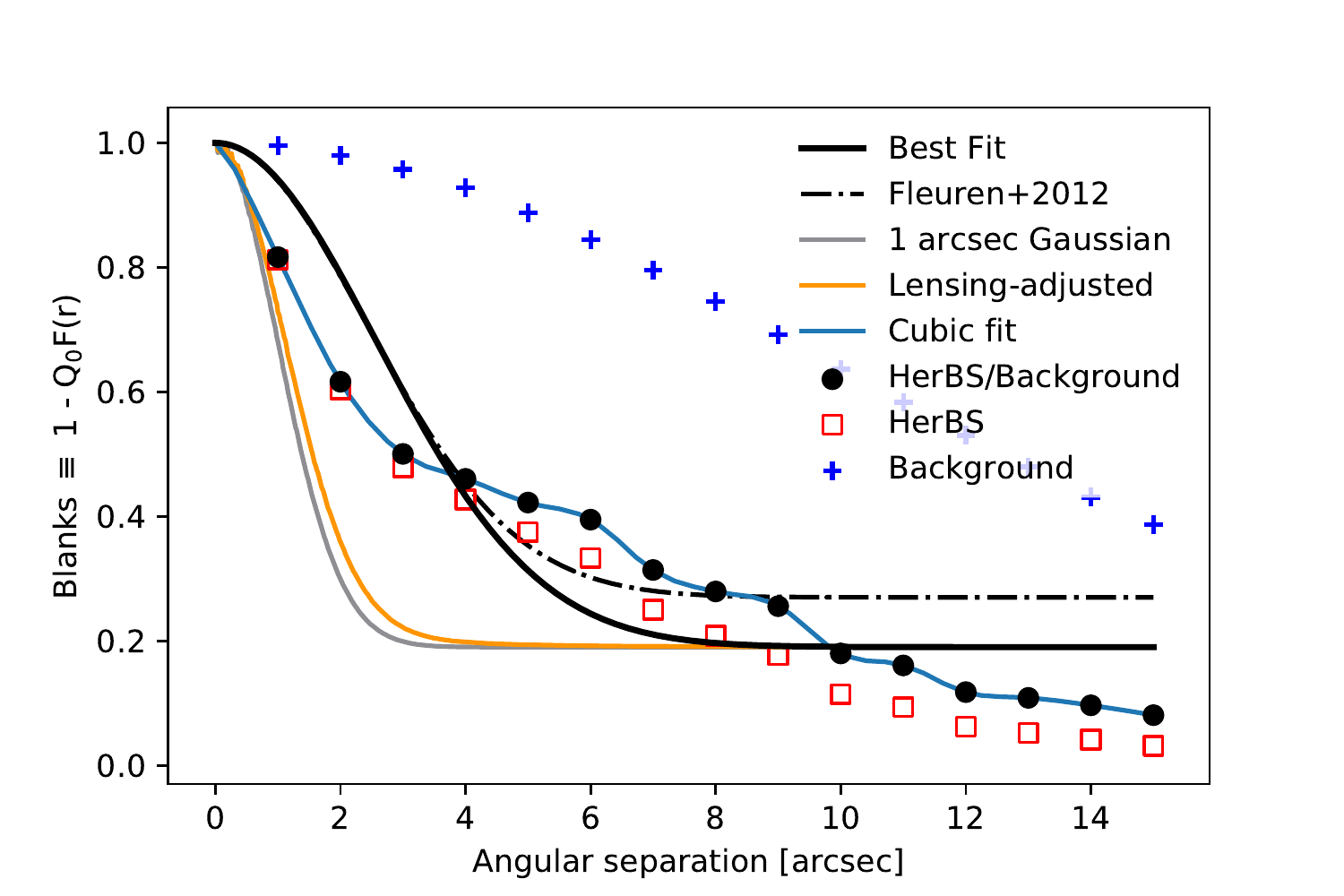}
\caption{B(r) (the 'blanks'), the fraction of positions without a VIKING galaxy within a radius r, plotted against radius. Random positions are shown with \textit{blue plusses} and the positions of HerBS sources are shown by \textit{red squares}. The \textit{black circles} show the result of dividing the two distributions, which corrects B(r) for the HerBS sources for unrelated VIKING galaxies falling within $r$ arcsec of the \textit{Herschel} position. The \textit{black line} shows the best-fitting Gaussian to this corrected distribution and the \textit{black dash-dotted line} shows the Gaussian fitted to the corrected distribution of \citet{Fleuren2012}. The poor fit-quality leads us to take $Q_0$ from the \textit{black circles} directly, at $\theta$ = 10, giving $Q_0$ = 0.82.
The (\textit{grey line} shows the form expected if lensing is not important and if the distribution is caused by astrometric errors with a FWHM of 1 arcsec, the value found by \citet{Bourne2016}.
The \textit{orange line} shows the distribution expected if we include both astrometric errors and the lensing offsets measured by \citet{Aris2018}. The large disagreement between the \textit{grey line} and the empirical results (\textit{black points}) shows that lensing is important. The large disagreement between the \textit{orange line} and the \textit{black points} shows that the lensing is occurring on a large angular scale and suggest it is not just the result of lensing by individual galaxies, \textcolor{referee}{since the \textit{orange line} shows the predicted behaviour of lensing by individual galaxies according to our Toy Model (Sec. \ref{sec:toymodel})}.}
\label{fig:Theta_VIKING}
\end{figure*}
In total, 98 HerBS sources are located within the VIKING images. Similarly, we distribute 10000 random positions over each VIKING image (3 $\times$ GAMA, and 1 $\times$ SGP), resulting in 40000 random positions in total, excluding regions within 10 arcseconds of HerBS sources. For each of these HerBS and random sources, we find all the VIKING galaxies within 15 arcsecond of the \textit{Herschel} position, and use this to calculate $Q_0$ as a function of radius, $Q_0(r)$, which is defined as the probability that a \textit{Herschel} source has a VIKING galaxy within $r$ arcseconds of the \textit{Herschel} position. \textcolor{referee}{Similar to \cite{Fleuren2012} and \cite{Bourne2016}, we used a limit of 15 arcseconds to estimate $Q_0(r)$, since $Q_0$ drops to zero beyond 15 arcseconds.}

We calculate $Q_0(r)$ using the method described by \cite{Fleuren2012}, which is not affected by clustering or multiple counterparts for \textit{Herschel} sources. For a given radius $r$ in arcseconds from (a) random postions and (b) \textit{Herschel} positions, we calculate the proportion of fields, B(r), for which there is no VIKING galaxy closer than $r$ to the central position.
If there were no galaxies unrelated to the \textit{Herschel} sources on the VIKING images, the fraction of \textit{Herschel} sources detected in VIKING would simply be equal to $Q_0(r)$ = 1 - B(r). Figure \ref{fig:Theta_VIKING} shows the blanks, B(r), for the random positions (\textit{blue plusses}) and for HerBS sources (\textit{red squares}). As the search radius $r$ increases, 
B(r) decreases steadily for the random positions but falls much more quickly for the \textit{Herschel} sources because of the near-IR counterparts to the \textit{Herschel} sources.

Before calculating $Q_0(r)$, B(r) for the HerBS sources has to be corrected for the galaxies that happen to fall within the radius $r$ by chance and are unrelated to the \textit{Herschel} source.
\cite{Fleuren2012} showed mathematically that one can correct for this by dividing B(r) for the HerBS positions by the B(r) for the random positions (\textit{red squares} / \textit{blue crosses}). The black points in Figure \ref{fig:Theta_VIKING} show B(r) corrected in this manner.

The true B(r) (\textit{black dots}) and the distribution of angular separations, $f(r)$, are directly related. We show this in the following two equations,
\begin{equation}
\textrm{B(r)} \equiv 1 - Q_0(r) = 1 - Q_0 F(r),
\label{eq:blanks}
\end{equation}
where $F(r)$ is the probability of finding a source between 0 and $r$. In other words, $F(r)$ is the integral of the distribution of angular separations, $f(r)$, along $r$,
\begin{equation}
F(r) = \int^r_0 2 \pi r^{\dagger} f(r^{\dagger})dr^{\dagger},
\label{eq:bigF}
\end{equation}
where $r^{\dagger}$ refers to the placeholder radius value for the integration.

If there was no lensing and the offsets between near-IR and \textit{Herschel} positions were the consequence of astrometric errors, the corrected distribution of B(r) (\textit{black points}) should follow the Gaussian distribution expected from the astrometric errors. 
Both \cite{Fleuren2012} and \cite{Bourne2016} made this assumption, which was a fair one because their samples consisted mostly of non-lensed sources.
However, we cannot assume a gaussian distribution, and we will simply take the value of $Q_0$ = 1 - B(r) at 10 arcseconds, giving $Q_0$ = 0.8200, although we note that B(r) seems to drop to 0 for greater angular separations.


In Figure \ref{fig:Theta_VIKING} we compare the background-corrected B(r) for the HerBS sources to the prediction of several models. Firstly, we fit the points with a Gaussian profile (\textit{black solid}), although it does not describe the points well, especially at larger angular separations. 
We further compare B(r) to the fit derived from the catalogue in \cite{Fleuren2012} (\textit{black dash-dot}), which shows a similar failure. The B(r) from \cite{Fleuren2012} is mostly derived from fainter sources with larger astrometric uncertainties. The \textit{grey} and \textit{orange lines} show the SDSS best-fit profiles for bright \textit{Herschel} sources, where sources with a S/N > 10 at 250$\mu$m have an astrometric error of less than 1 arcsecond \citep{Bourne2016}, which also holds true for almost all HerBS sources in this comparison. The \textit{orange line} is the lensing-adjusted value from the SDSS analysis \textcolor{referee}{(Sec. \ref{sec:toymodel})}. None of these profiles represent the data adequately. 

\subsubsection{Estimation of lensing-adjusted f(r)}
\label{sec:newfofr}
Figure~\ref{fig:Theta_VIKING} shows that all the analytic models based on the assumption that the form of B(r) is only the result of astrometric errors fail.

Instead, we will derive the distribution of angular separations from the B(r) directly. We do this by taking the derivative of equation \ref{eq:bigF},
\begin{equation}
\frac{dF(r)}{dr} = \frac{d}{dr}\int^r_0 2 \pi r^{\dagger} f(r^{\dagger})dr^{\dagger} = \left[2 \pi r^{\dagger} f(r^{\dagger})\right]_0^r = 2 \pi r f(r).
\end{equation}
Re-arranging, and including equation \ref{eq:blanks}, leads to
\begin{equation}
f(r) = - \frac{1}{2\pi Q_0 r}\frac{d(\textrm{B}(r))}{dr},
\label{eq:findnewdistribution}
\end{equation}
where it is important to note the potential but nonphysical instability at $r$ = 0.
We take $Q_0$ as being the value we estimate at r~=~10 arcsec, which is equivalent to forcing the density profile,  $f(r)$, to be be equal to 0 for any $r$ $>$ 10 arcsec.
We find the derivative of B(r) by a cubic interpolation routine in python, which fits the values between the points continuously. \textcolor{referee}{This fit is shown in Figure \ref{fig:Theta_VIKING} as the \textit{cubic fit} between the \textit{black points}.}

\begin{figure}
\centering
\includegraphics[width=\linewidth]{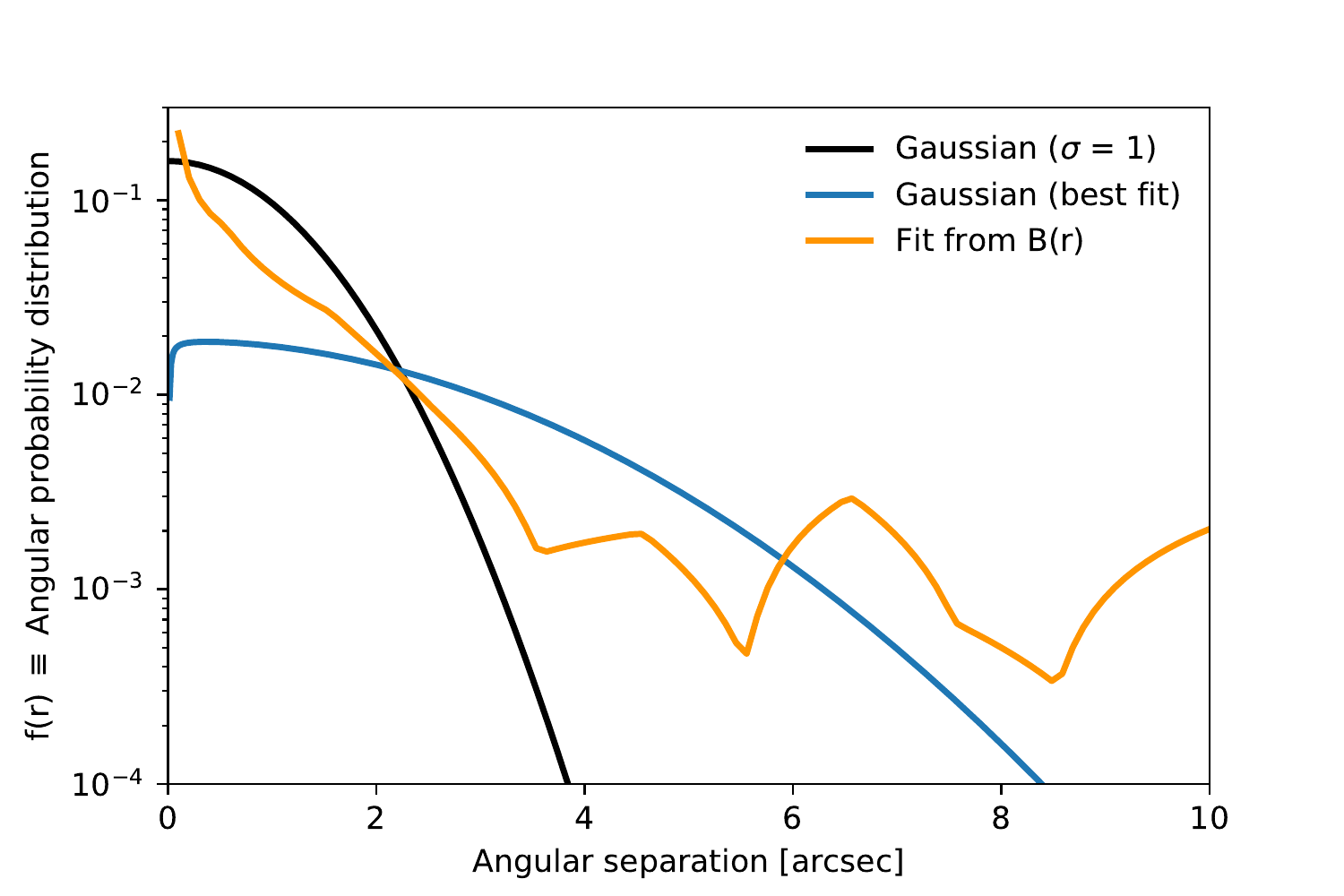}
\caption{The angular separation distribution, $f(r)$, is plotted against the radius. Using equation \ref{eq:findnewdistribution}, we derive a new angular distribution for the HerBS sources (\textit{orange}), which appears different to the Gaussian profile we have seen for non-lensed sources. We plot both a Gaussian distribution with $\sigma$ = 1 arcsecond (\textit{black line}), which is the expected error in our astrometry, and a Gaussian distribution which was fitted to our B(r).}
\label{fig:angdistVIK}
\end{figure}

We show the angular probability distribution, $f(r)$, in Figure~\ref{fig:angdistVIK}. The \textit{orange line} shows the $f(r)$ derived from B(r). The \textit{black line} shows the Gaussian that one expects if lensing is not important and the form of  $f(r)$ is the result of astrometric errors. The \textit{blue line} shows the Gaussian fit that we find from a fit of B(r). The difference between the \textit{black} and \textit{orange} lines shows clearly that some process is at work in our sample beyond simple astrometric errors. Since we expect many of our sources to be lensed \citep{GN2012,GN2014,GN2017,Bourne2014,Negrello2016}, 
we conclude that the extended form of $f(r)$ is due to gravitational lensing. 

\subsubsection{Deriving the magnitude distributions}
We calculate the magnitude distribution of the genuine counterparts, $q(m)$, by comparing the magnitude distributions of the galaxies within 10 arcseconds of the Herschel positions ($n_{\textrm{total}}$) and within 10 arcsec of the random positions ($n_{\textrm{back}}$). We take a 10 arcsecond search radius, similar to both \cite{Fleuren2012} and \cite{Bourne2016}. We use the following relationship to estimate the magnitude distribution of the galaxies associated with the Herschel sources.
\begin{equation}
n_{\textrm{real}}(m) = \frac{n_{\textrm{total}}}{\textrm{Area}_{\textrm{total}}} - \frac{n_{\textrm{back}}}{\textrm{Area}_{\textrm{back}}}.
\end{equation}
Then, we apply a normalization to ensure that the integral of $q(m)$ is equal to the probability that a source is visible in the VIKING fields, $Q_0$,
\begin{equation}
q(m) = Q_0 \frac{n_{\textrm{real}}(m)}{\int_{-\infty}^{\infty} n_{\textrm{real}}(m) dm}.
\label{eq:calcqm}
\end{equation}
The background surface distribution, $n(m)$, is given by equation~\ref{eq:nback}, repeated here,
\begin{equation}
n(m) = \frac{n_{\textrm{back}}(m)}{\textrm{Area}}.
\end{equation}
Here Area refers to the total area probed by all the random positions, thus equal to the number of random positions times $\pi \times 10 \times 10$ square arcseconds.

\begin{figure}
\centering
\includegraphics[width=\linewidth]{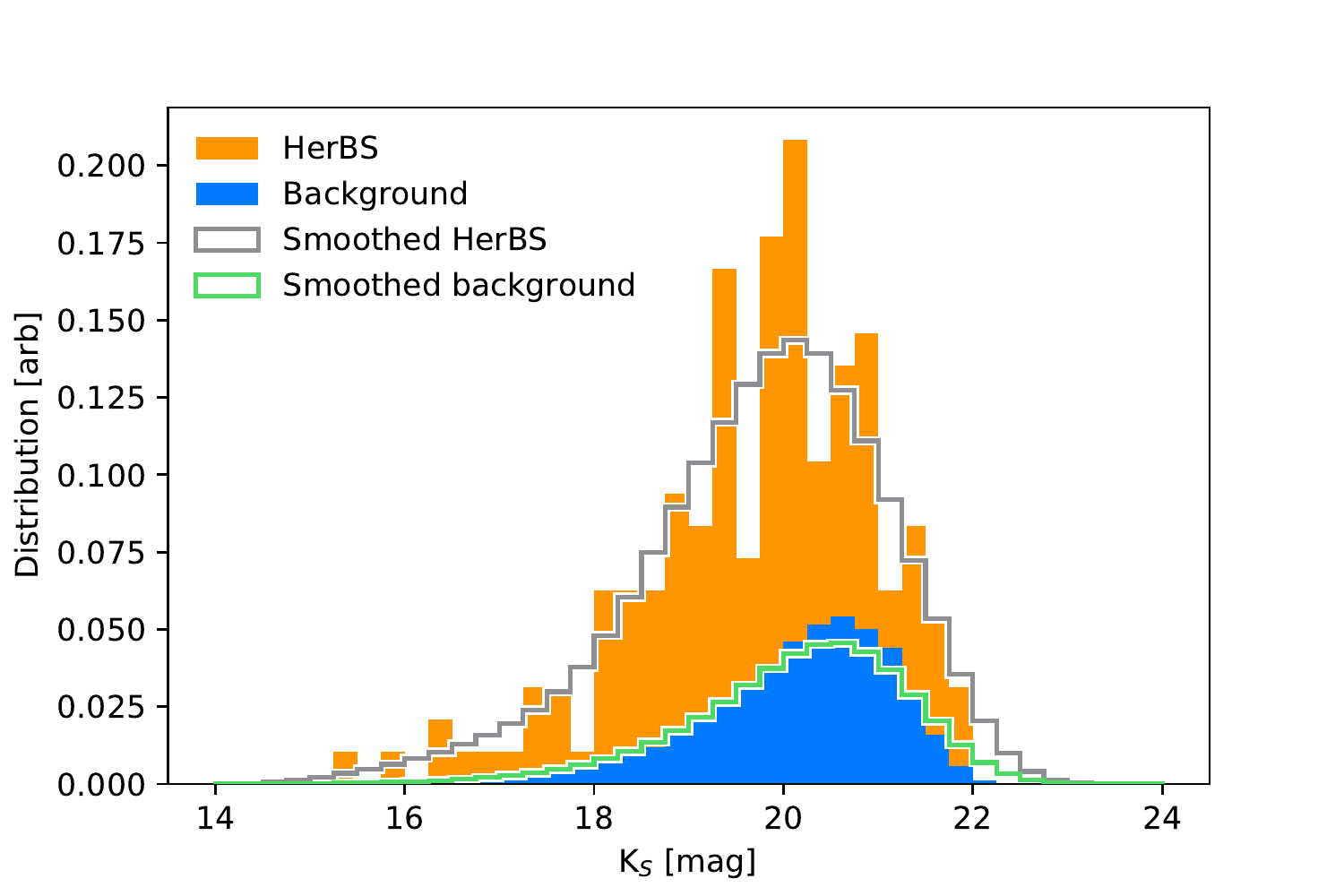}
\caption{The \textit{orange histogram} shows the magnitude distribution within 10 arcseconds of HerBS sources, and the \textit{solid blue histogram} shows the background magnitude distribution within 10 arcseconds of random positions. The small number of sources contributing to the HerBS magnitude distribution would give a noisy estimation of the true HerBS magnitude distribution, $n_{\textrm{real}}(m)$ (\textit{black dashed line}). We smooth the histograms using a Gaussian spread with a FWHM of 0.5 magnitudes, which gives the \textit{grey histogram} and \textit{green histogram} for the HerBS and background magnitude distributions, respectively.}
\label{fig:hists_VIKING}
\end{figure}
Whereas \cite{Fleuren2012} and \cite{Bourne2016} have thousands of \textit{Herschel} sources to estimate their probability distributions, we have less than a hundred. This can be seen in Figure \ref{fig:hists_VIKING}, which shows the magnitude distribution of both the HerBS and random positions. The \textit{orange histogram} shows the magnitude distribution for the HerBS positions, and the \textit{solid blue histogram} shows the magnitude distribution for the random positions.

If we were to simply use these distributions, it will result in a noisy estimation of the true HerBS magnitude distribution ($n_{\textrm{real}}(m)$) due to low-number statistics. Similarly, the small number of data points creates a non-continuous magnitude distribution, which is an inconvenience for a successful implementation of the statistical method. 

We apply a simple gaussian smoothing to the two histograms, which should decrease bin-to-bin variation, exposing the global trend of $n_{\textrm{real}}(m)$, shown by the \textit{grey histogram} and \textit{green histogram} for the HerBS and background magnitude distributions, respectively. We choose a gaussian with a width of 0.5 magnitude, narrow enough to not vary the distribution, but wide enough to ensure the low-magnitude region is adequately smoothed.

\begin{figure}
\centering
\includegraphics[width=\linewidth]{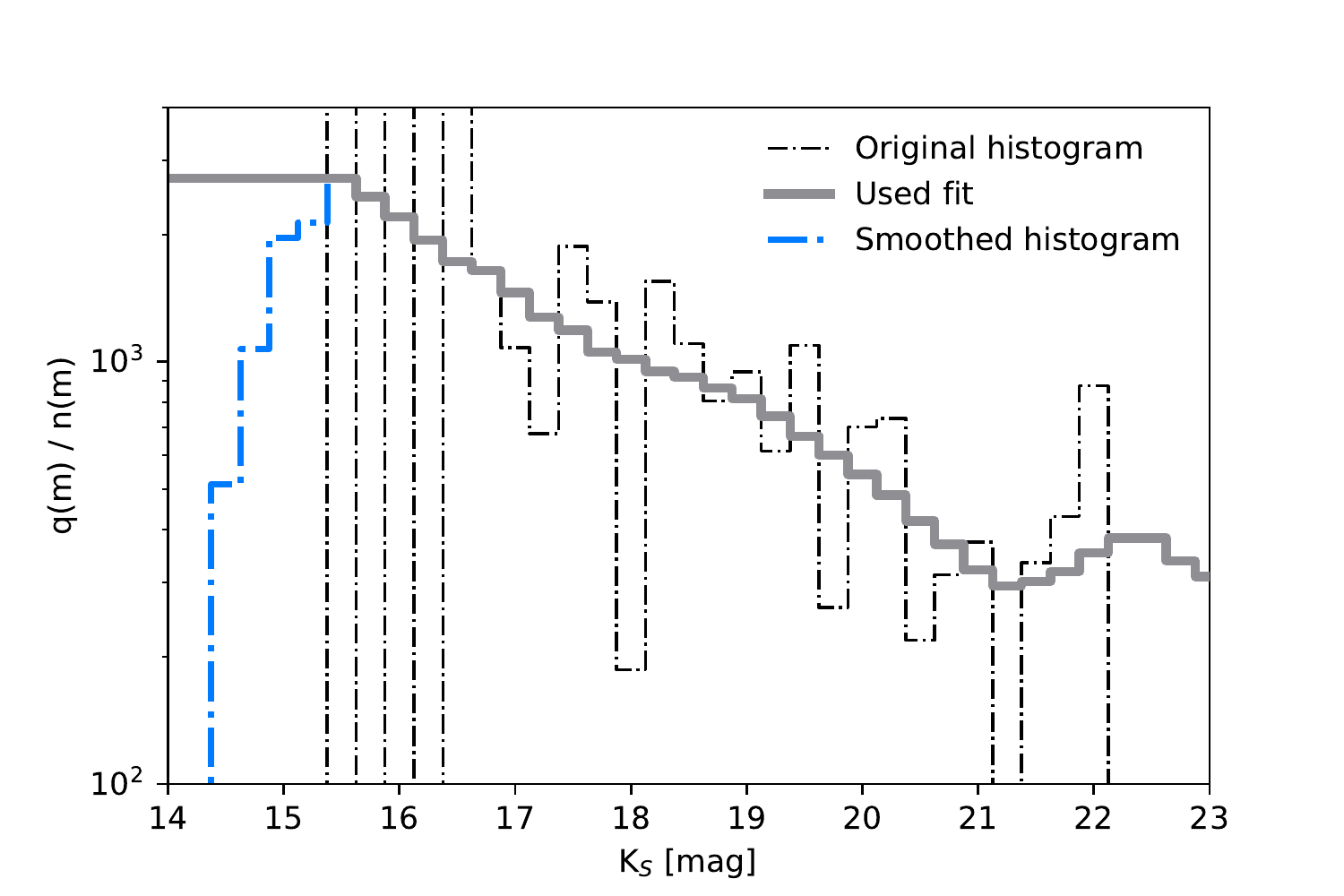}
\caption{The expected number of genuine counterparts, $q(m)$, divided by the background VIKING interloper, $n(m)$, are estimated by first smoothing the histograms with a Gaussian distribution. The likelihood value calculated for each counterpart is the multiplication the value in Figure \ref{fig:angdistVIK} at the radial offset by value inside this graph at the magnitude of the source. 
The \textit{blue histogram} is calculated by smoothing the magnitude distributions by a gaussian with 0.5 magnitude, narrow enough to not vary the distribution, but wide enough to ensure the low-magnitude region is adequately smoothed. The noise in the original source distribution (\textit{black line}) shows the need for smoothing. The low values of $q(m)/n(m)$ for bright sources appears unphysical, which we circumvent by fixing the value, and use the \textit{grey histogram} smoothing to derive the likelihood.}
\label{fig:qn_VIKING}
\end{figure}

In Figure~\ref{fig:qn_VIKING}, we divide the genuine counterparts probability distribution, $q(m)$, by the background surface density of VIKING galaxies, $n(m)$. We calculate $q(m)$ from equation \ref{eq:calcqm}.
We show the smoothed histogram as the \textit{blue histogram}. 
Because of low number statistics at bright magnitudes, we used a constant value for q(m)/n(m) at m $< 16$, the same approach used by \cite{Fleuren2012} and \cite{Bourne2016}.

\cite{Fleuren2012} found typical values of $q(m)/n(m)$ ranging from 10000 at K$_{\rm{S}}$ = 15 mag to $\sim$ 200 at K$_{\rm{S}}$ = 21 mag. Both our methods give values within these ranges, but the distribution of $q(m)/n(m)$ has a less steep dependence on magnitude than that found by \cite{Fleuren2012}.

\subsection{Results}
We summarize the results of the reliability analysis in Table~\ref{tab:VIKreliabilities}. We find, over the entire VIKING fields, 56 sources have counterparts with R $>$ 0.8, equal to 57\% of sources. This is more than we found for the SDSS counterpart search in Section \ref{sec:sdss}. There we found 31 out of 121 potential foreground galaxies (25\%) directly from the catalogues of \cite{Bourne2016}, which we were able to increase to 32.5 or 41 likely SDSS counterparts (27\% and 35\%, respectively) when we accounted for gravitational lensing, depending on the method we used. Our $Q_0$ value at $r$ = 10 arcsec, 0.82, implies that 82\% of HerBS sources actually have counterparts, although we are only able to identify the counterparts for 57\% of the individual sources (the other sources must have offsets or magnitudes that lead to values of R $<$ 0.8). We also note that our choice of 10 arcsec as the radius at which to calculate $Q_0$ was chosen because it had been used in previous work. $Q_0$ increases with radius and reaches close to 100\% at $r$ = 15 arcsec, suggesting that all HerBS sources have some statistically associated foreground galaxy present on the VIKING images.
\begin{table}
    \caption{VIKING reliabilities}  \label{tab:VIKreliabilities}
    \begin{tabular}{lccc} \hline
          R & < 0.8  & > 0.8 & All \\ \hline
        \textbf{SGP} 		&  	6	&	19	& 28  		\\
        \textbf{GAMA09} 	&  	13	&	9	& 21  		\\
        \textbf{GAMA12} 	&  	9	&	16	& 26  		\\
        \textbf{GAMA15} 	&  	10	&	12	& 23  		\\
        \textbf{Total} 		&  	42	&	56	& 98  		\\
        \hline
    \end{tabular}
\textbf{\\Notes:} Reading from the left, the columns are: Column 1 - the field; column 2 - sources with reliabilities less than 0.8; column 3 - sources with reliabilities greater than 0.8; column 4 - the total number of sources in each field.
\end{table}

\begin{figure*}
\centering
\includegraphics[height=0.9\textheight]{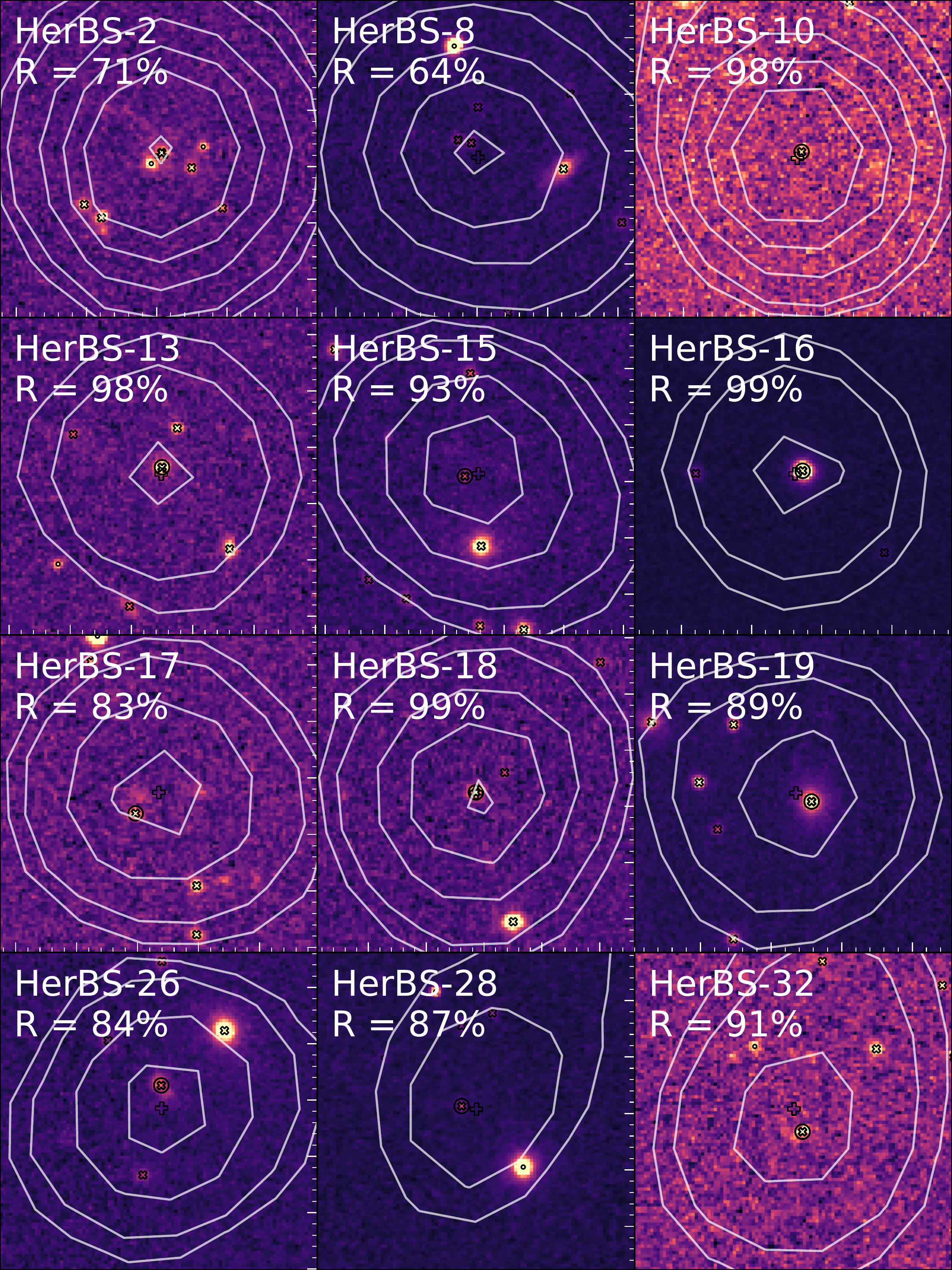}
\caption{This figure is the first of twelve cutouts of HerBS sources in the VIKING fields. The 30 by 30 arcsecond VIKING image is centred on the \textit{Herschel} 250 $\mu$m position, which is indicated by a plus, and the contours are placed at 30, 50, 100, 150, 200 and 300 mJy. All VIKING-extracted sources with J - K$_{\rm{S}}$ > 0 (non-stars) are indicated with a cross, and the most likely counterpart has a circle placed around it. Assumed stars are indicated with a small circle. We mention the reliability in terms of a percentage.}
\label{fig:VIKstamps}
\end{figure*}

We have shown pictures of small areas of the VIKING images around the Herschel positions for the first 12 of the HerBS sources in Figure \ref{fig:VIKstamps} and for the remaining 86 in \textcolor{referee}{the online supplementary material}. Each panel consists of a 30 by 30 arcsecond cutout of the K$_{\rm{S}}$-band image, centered on the 250 $\mu$m \textit{Herschel} position, which is indicated by a plus. We show the VIKING objects with J - K$_{\rm{S}}$ > 0 (crosses), where we highlight the 
VIKING object that is most likely to be associated with the
\textit{Herschel} source, if present, with a circle. All VIKING objects with J - K$_{\rm{S}}$ < 0 have been marked with a small circle, and are assumed to be stars. The white lines indicate the 250 $\mu$m contour lines, which we choose, as this is the flux \textcolor{referee}{density} at which the position is determined by \cite{Valiante2016}. 

We tabulate all sources in the Appendix Table \ref{tab:totalSources}, where we list the sub-mm and near-infrared redshift (see Section \ref{sec:vik_z_phot}), the reliability, K$_S$ photometry, angular offset and VIKING position for each HerBS source. We indicate the sources with near-infrared photometric redshifts from \cite{Wright2018} with a $\dagger$.

\section{Discussion}
\label{sec:discussion}
In Section \ref{sec:vik_z_phot} we check the possibility that the VIKING images are sensitive enough to see the galaxies producing the sub-mm emission rather than the foreground lenses, using photometric redshifts. In Section \ref{sec:VIKING_opt_nir}, we estimate the fraction of the HerBS sources that are lensed. In Section \ref{sec:discCompare}, we compare our lensing results with previous results. In Section \ref{sec:discAng}, we explore why we see a different angular distribution than expected from a lensing model for individual galaxies. Finally, in Section \ref{sec:discExtended}, we expand our foreground search to the entire H-ATLAS catalogue \citep{Valiante2016,Maddox2018}.

\subsection{VIKING and HerBS redshift separation}
\label{sec:vik_z_phot}
In order to ensure the VIKING galaxy is not the background sub-mm source itself, we compare the photometric redshift of the \textit{Herschel} source to the photometric redshift derived from VIKING flux \textcolor{referee}{densities}. We use the photometric redshifts given in \cite{Bakx2018} for the background sub-mm sources, which are derived by fitting a two-temperature modified black-body to \textit{Herschel}/SPIRE 250, 350 and 500 $\mu$m photometry. If available, JCMT/SCUBA-2 850 $\mu$m flux \textcolor{referee}{densities} were used to improve the photometric redshift estimate. We have used, where available, the photometric redshifts from \cite{Wright2018} for the redshifts of the VIKING galaxies. \cite{Wright2018} used both optical KiDS and near-infrared VIKING photometry to calculate the photometric redshifts in this catalogue. If the source is not covered in this catalogue, we estimate the photometric redshift by applying the Eazy-photometric package to the five-band VIKING photometry \citep{Brammer2008}.
\begin{figure}
\centering
\includegraphics[width=\linewidth]{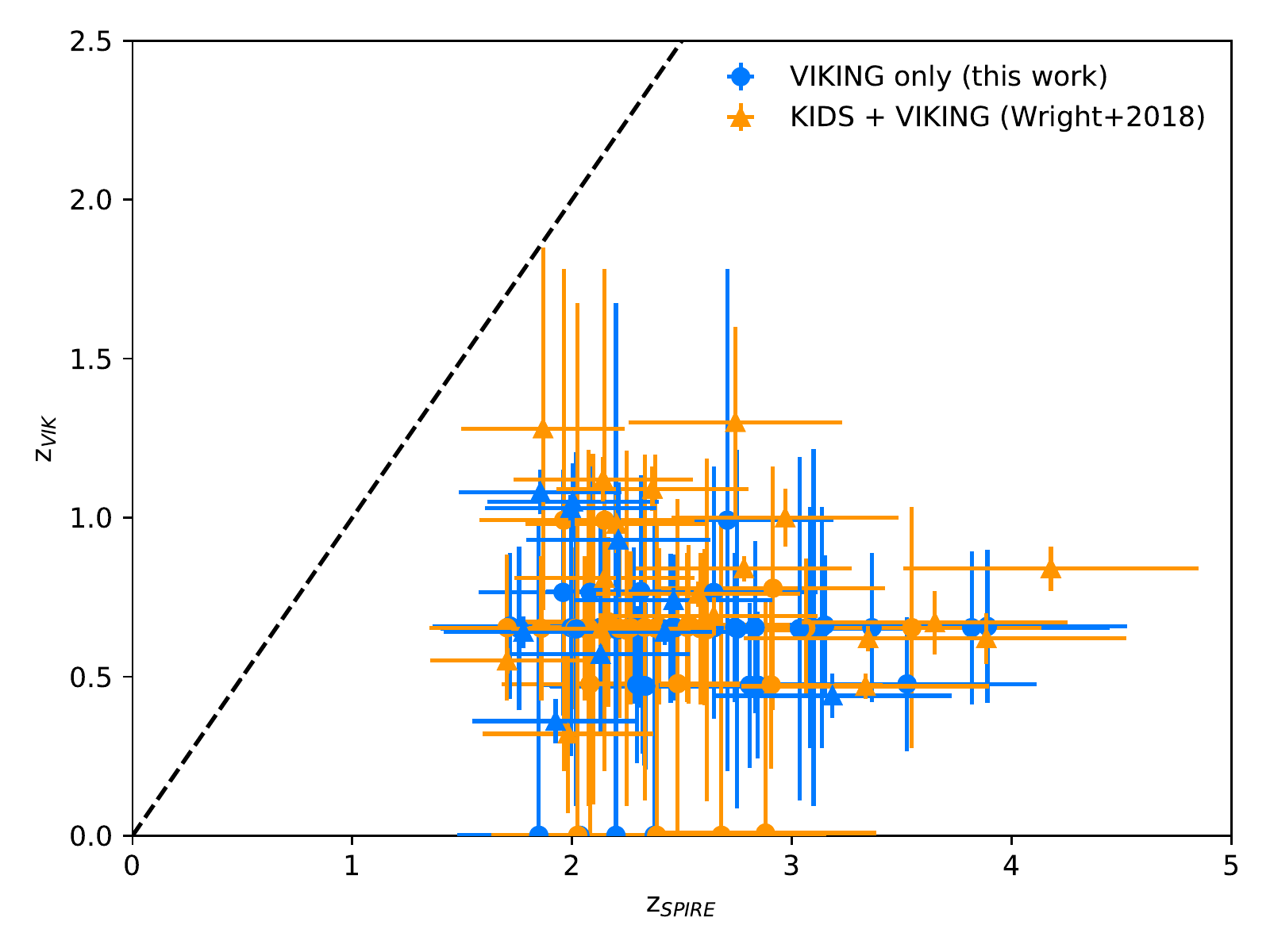}
\caption{The photometric redshift estimates from \textit{Herschel}/SPIRE-based redshifts are shown against the photometric redshifts from VIKING flux \textcolor{referee}{densities} for all sources with a reliability greater than 0.8. We use the sub-mm redshifts from \protect\cite{Bakx2018}. The VIKING-based photometric redshifts are, where available, extracted from \protect\cite{Wright2018}, which uses VIKING and KiDS photometry. If this is not available, we use the Eazy package to calculate the photometric redshifts using the VIKING flux \textcolor{referee}{densities} extracted in this paper. A single source is located close to the y = x line, and has a 10\% chance to be at the same redshift, but all other sources are less than 1\% likely to be the same source.}
\label{fig:zspec_comparison}
\end{figure}

In Figure \ref{fig:zspec_comparison} we have plotted the photometric redshifts of the \textit{Herschel} sources against the photometric redshifts of the VIKING galaxies. It suggests all our VIKING-observed sources are at lower redshift than our \textit{Herschel}-selected sources. 
We calculate the probability that both redshifts are of the same object by assuming a Gaussian probability distribution in both the VIKING and sub-mm photometric redshift errors. The Eazy package provides the near-infrared photometric redshift errors, and for the sub-mm photometric redshift errors we use $\delta z / (1+z)$ = 0.13 from \cite{Bakx2018}. 
For the object shown by the top-left point in the figure, there is a probability of $\sim$10\% that the two redshifts are the same.
All the other sources have a probability of less than 1\%, down to 10$^{-4}$\%, to be at the same redshift.
Hence, we feel confident that we are observing foreground sources in the VIKING survey, of which we expect most to be lensing galaxies. 

\subsection{The lensing nature of optical and near-IR counterparts}
\label{sec:VIKING_opt_nir}
\begin{figure*}
\centering
\includegraphics[width=\linewidth]{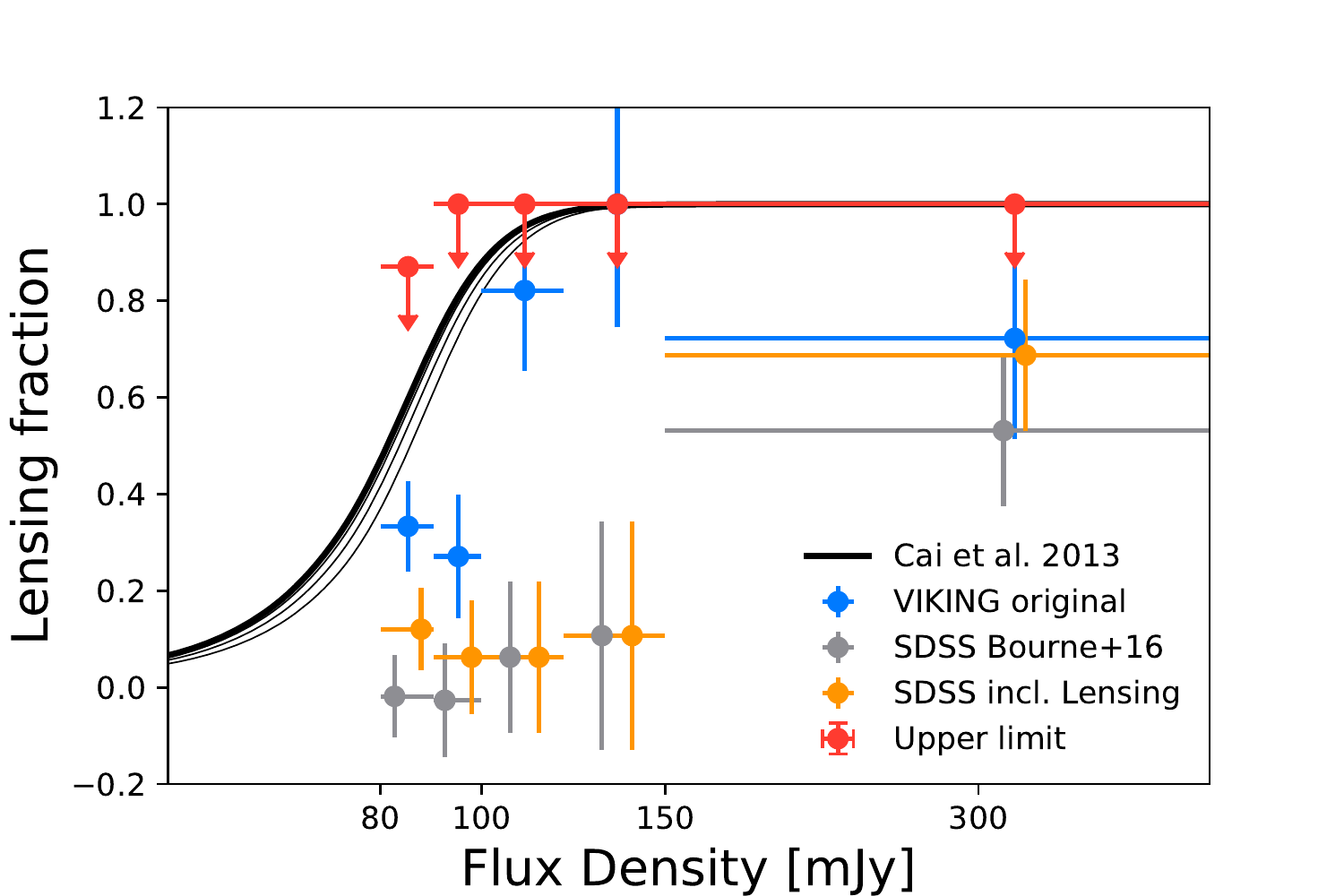}
\caption{The \textit{blue points} show our estimates of the fraction of HerBS sources with reliable (R $>$ 0.8) counterparts on the VIKING near-IR images plotted against 500-micron flux density. The \textit{grey points} show estimates of this fraction from the SDSS-based catalogues of \citet{Bourne2016}. The \textit{orange points} show the estimates from the SDSS-based catalogues corrected using the model described in the \ref{sec:toymodel} for the lenses that are too far from the \textit{Herschel} position or or have too faint a magnitude to give R $>$ 0.8. The \textit{red points} are upper limits on this fraction using the method described in the text. The four lines show the lensing fraction estimates from \citet{Cai2013}, where the bold line shows the lensing fraction with an assumed maximal magnification assumption of 30, and the thin lines correspond to 20, 15 and 10.}
\label{fig:lensFrac}
\end{figure*}

\noindent
Figure \ref{fig:lensFrac} shows the lensing fraction, our estimate of the fraction of lensed sources over the total number of sources. We compare this lensing fraction to the model of \cite{Cai2013}. However, our lensing fraction in flux \textcolor{referee}{density} bin i is not simply equal to the fraction of counterparts with a reliability greater than a certain threshold ($\rm F_{R>R_{thresh},i}$). Instead, this fraction of counterparts also includes a certain amount of sources that are included as false positives. For example, if the true lensing fraction ($\rm F_{{lens,i}}$) were 0.5 in a given flux density bin, 50\% of sources do not have a foreground galaxy. Of those non-lensed sources, we can expect a fraction of 1 - $\rm R_{thresh}$ to be included as false positives. This effect thus becomes particularly noticeable at low lensing fractions, where most apparent counterparts are likely to be false positives. In the most extreme case, if there are no lensed galaxies (i.e. the lensing fraction is 0), the fraction of sources with a counterpart greater than R$_{\rm thresh}$ flattens out to 1 - R$_{\rm thresh}$. To compensate for this flattening effect, we calculate the fraction of gravitational lenses from the fraction of sources with observed counterparts in flux density bin i, using
\begin{eqnarray}
\rm F_{{lens,i}} &=& \rm \left(F_{R>R_{thresh},i}\right) - (1 - R_{thresh}) (1 - F_{{lens,i}}), \\ 
\rm F_{{lens,i}} &=& \rm \frac{1}{R_{thresh}}\left(F_{R>R_{thresh},i} - 1\right) + 1. \label{eq:RatioToLF}
\end{eqnarray}

In Figure \ref{fig:lensFrac}, $\rm R_{thresh}$ is set to 0.8, and the error bars are calculated by dividing one by the square root of the number of sources in each bin divided by (1 - R$_{\rm thresh}$). The blue points show the lensing fraction for the HerBS sources covered by VIKING. The grey points show the lensing fraction of HerBS sources with SDSS counterparts, using the results from \cite{Bourne2016} and \cite{Furlanetto2018}, and the orange points show the lensing fraction of HerBS sources with SDSS counterparts calculated with their lensing-corrected reliabilities using the model described in the Section \ref{sec:sdss}. The upper limits (\textit{red}) are from the fraction of HerBS sources without any VIKING galaxy visible at $r$ $<$ 10 arcsec. By subtracting this fraction from 1 we obtain an upper limit on the fraction of HerBS sources with lenses visible on the VIKING images. 
We plot four realizations of the lensing fraction from the galaxy evolution model by \cite{Cai2013}, where the thick black line has a maximum magnification, $\mu_{\textrm{max}}$, = 30. The other three, thinner lines correspond to the realizations with 20, 15 and 10 as their maximum magnification (M. Negrello, private communication).

The scatter on the calculated values is large, but an increase in the lensing fraction is seen for VIKING galaxies with increasing 500 $\mu$m flux \textcolor{referee}{densities}. The galaxy evolution model suggests a significant fraction of lenses at lower flux \textcolor{referee}{densities}. At the high flux \textcolor{referee}{densities}, 500$\mu$m $>$ 100mJy, we find a lensing fraction of nearly 1, as expected from \cite{Negrello2010}. At lower flux \textcolor{referee}{densities}, however, we find that the lensing fraction drops quickly to around 30 to 40\%, either suggesting a faster drop-off in lenses than expected from the model by \cite{Cai2013}, or that even the VIKING survey is not able to find all foreground lensing galaxies.
The lensing fraction for 500$\mu$m $<$ 50mJy sources is around 0. This low lensing fraction could suggest that the majority of the counterparts are false positives. In order to fully account for all lensed sources, therefore, we recommend using near-infrared, deep observations to look for the foreground lensing sources. In the future, this could be done with the SHARKS project, which covers the SGP and GAMA 12 and 15 fields, and achieves a four times deeper K$_{\rm S}$ depth than VIKING (5$\sigma$ depth: 22.7 mag$_{\rm AB}$, P.I. H. Dannerbauer \footnote{\url{https://www.eso.org/sci/observing/PublicSurveys/sciencePublicSurveys.html}}).

\subsection{Comparison to other lensing information}
\label{sec:discCompare}
\subsubsection{Confirmed lenses in SDSS and VIKING}
We compare our SDSS and VIKING results to the lensing classifications in \cite{Negrello2016}. The SDSS analysis includes 37 of the sources of \cite{Negrello2016}. Out of these 37 sources, 14 are classified as confirmed gravitational lenses ('A'-category). The SDSS reliability from our revised calculations (Section \ref{sec:sdss}) is R > 0.8 for 11 sources, is R < 0.8 for one source, and two sources do not have any nearby SDSS galaxies. Four of the 37 sources are in the 'B'-category, likely lensed sources. Three of these sources have R > 0.8, and one source has R < 0.8. 18 out of the 37 sources are in the 'C'-category, unidentified sources, one source has R > 0.8, six sources have R < 0.8 and 11 sources do not have any nearby SDSS galaxies.

The VIKING analysis includes 28 sources also documented in \cite{Negrello2016}. Of the five sources with 'A'-category, confirmed lenses, four sources have R $>$ 0.8, and one source has R < 0.8. Two sources are in the 'B'-category, likely lensed sources, and have a reliability greater than 0.8. Of the 20 sources with 'C'-category, unidentified sources, 18 sources have R > 0.8, one source has R < 0.8, and one source does not have a lens identification nearby.

The only source in \cite{Negrello2016} with a 'D'-category (HerBS-8; \citealt{Ivison2013}), confirmed to be not strongly lensed. This is consistent with the fact that we didn't find a galaxy on either the SDSS or VIKING with R $>$ 0.8. 

\subsubsection{Comparison to SHALOS}
\begin{figure}
\centering
\includegraphics[width=\linewidth]{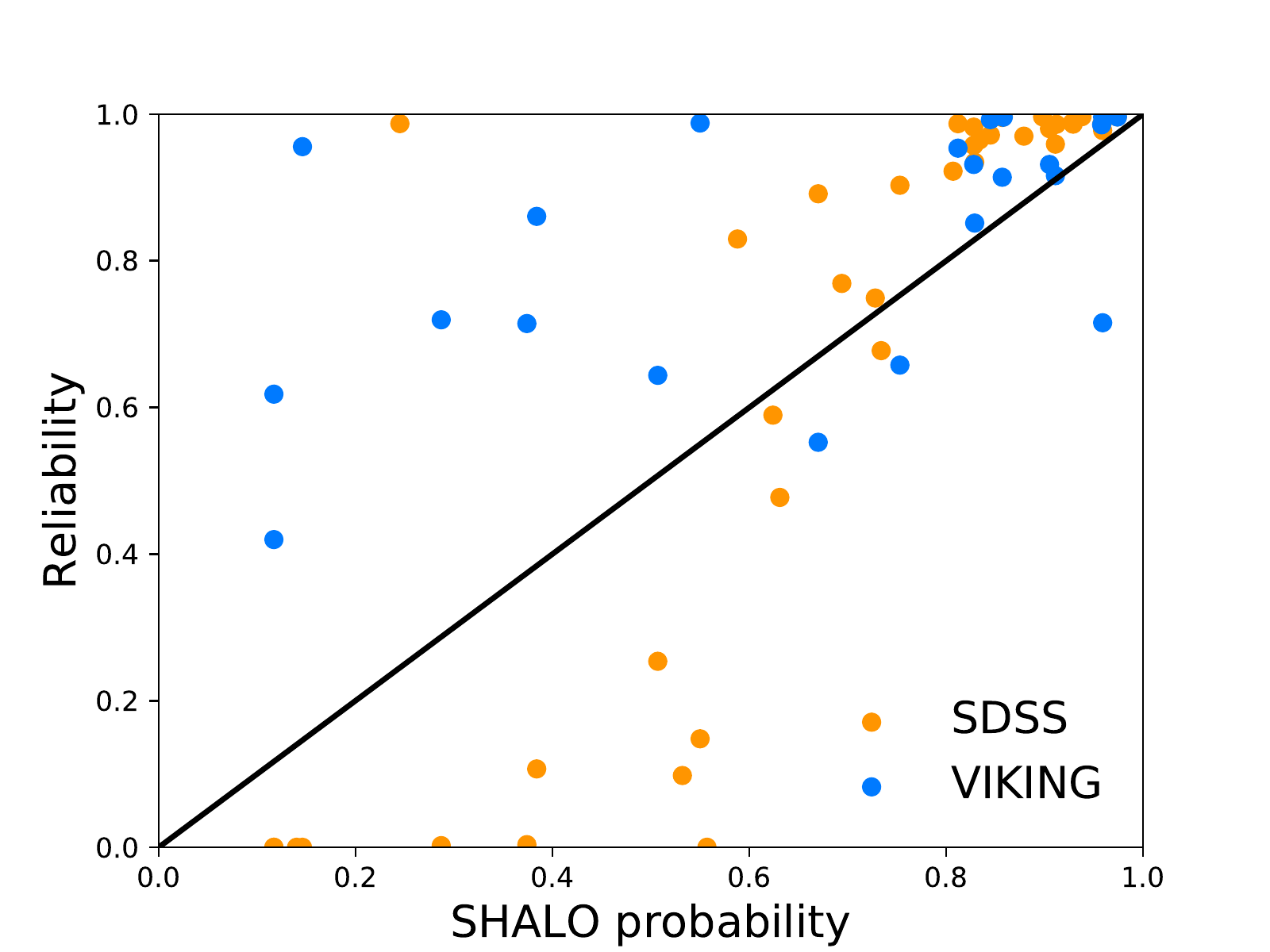}
\caption{We show the Reliability in both SDSS (\textit{orange}) and VIKING (\textit{blue}) against the lensing probability according to the catalogue from \citet{GN2019}. More than half of the counterparts agree with the analysis of \citet{GN2019}, although we do find disagreement, where our reliability finds lower lensing probabilities for the SDSS galaxies, and higher lensing probabilities for the VIKING galaxies.}
\label{fig:SHALOS}
\end{figure}
We compare our updated SDSS and VIKING results to the catalogue from \cite{GN2019} in Figure \ref{fig:SHALOS}. They provide a list of lensing probabilities from a probabilistic estimator that uses the optical and sub-mm flux \textcolor{referee}{densities}, the angular distance between the optical and sub-mm positions, and the estimated redshifts of the sub-mm and optical sources to look for galaxies on the SDSS that are lensing \textit{Herschel} sources. 

In total, 42 HerBS sources with an SDSS reliability, and 23 HerBS sources with a VIKING reliability are also in the SHALOS catalogue of \cite{GN2019}. 
\begin{table}
    \caption{SHALOS comparison to SDSS / VIKING}  \label{tab:SHALOS}
    \begin{tabular}{lccc} \hline
        \textbf{SHALOS}     &   R > 0.8         & R < 0.8   & All \\ \hline
        \textbf{p $>$ 0.8}	&  	22 / 12	        &	0 / 1   & 22 / 13  	\\
        \textbf{p $<$ 0.8} 	&  	4 / 3	        &	16 / 7  & 20 / 10  	\\
        \textbf{Total} 		&  	26 / 15	        &	16 / 8  & 42 / 23  	\\
        \hline
    \end{tabular}
\end{table}
We quantify this comparison between the SHALOS and SDSS and VIKING reliability in Table \ref{tab:SHALOS}. 
We find that our method, for both SDSS and VIKING, agrees with the SHALOS method quite well. Most sources with SHALOS probabilities greater than 0.8 have SDSS and VIKING reliabilities greater than 0.8. The same holds for most sources with SHALOS probabilities less than 0.8, which typically have SDSS and VIKING reliabilites less than 0.8. For sources with SDSS reliabilities less than 0.7, the SDSS reliability is lower than the SHALOS probability. We note the opposite for sources with VIKING reliabilities less than 0.7, where the reliability is larger than the SHALOS probability. This variety between the SHALOS method and our methods is because each method assumes a different statistical contribution of the angular offset between the foreground and background sources.

We know this, because the SHALOS catalogue by \cite{GN2019} specifies the individual components (i.e. probabilities) of the probabilistic estimator (e.g. magnitude, angular separation, ...). For all but one source, all individual probabilities of SHALOS are close to 1, except the probability associated with the angular offset between the foreground and background source. Therefore, the dominant factor determining the SHALOS probability is the angular separation. \cite{GN2019} assumes a single value for the global astrometric RMS precision of 2.4 arcsec, whilst our SDSS analysis assumes a 250$\mu$m flux \textcolor{referee}{density}-dependent precision (typically around 1 arcsec; \citealt{Bourne2016}), and our VIKING analysis assumes an angular distribution that accounts for a more extended angular distribution (see Sec. \ref{sec:newfofr}). 

\subsection{Large angular offsets}
\label{sec:discAng}
Both our analysis of the optical SDSS and near-infrared VIKING galaxies find angular offsets larger than expected. In Figure \ref{fig:MCs}, we find a disagreement between the angular offsets between the SDSS galaxies and HerBS sources, and the offsets predicted by toy model (Sec. \ref{sec:toymodel}) of the angular distribution that takes into account the additional offset generated by gravitational lensing. This toy model is based on actual ALMA observations of 15 lensed sources \citep{Aris2018}. This discrepancy suggests a lack of galaxies with short angular separation distances between \textit{Herschel} and SDSS positions. We calculate the number of missed SDSS candidates using two methods (Sec. \ref{sec:GravLensingOnSDSS}). Each method results in very different estimates on the number of missed lenses (e.g. 1.5 or 10 missed gravitational lenses). The first method calculates the number of missed lenses from only the SDSS galaxies that have a reliability greater than 0.8. The second method recalculates the reliabilities for the SDSS sample using our toy model angular offset distribution adjusted for lensing. The first method thus assumes that the SDSS galaxies with R $>$ 0.8 are a good representation for the full sample of lensed sources. This is not necessarily true, as we can see in Figure \ref{fig:ActVsMax}. Many galaxies have reliabilities close to, but below R $<$ 0.8. When instead we apply the toy model to all sources, regardless of reliabilities, (i.e. the second method) we see in Figure \ref{fig:NewMax} that most of these galaxies are shifted into R $>$ 0.8. This causes a large increased in the estimate of missed lenses.

The VIKING B(r), the fraction of sources without any nearby counterpart for a radius $r$, shown in Figure \ref{fig:Theta_VIKING} does not agree with the expected angular distribution (i.e. the toy model). The blanks suggest that the angular separation distribution extends out to much larger angular scales. We account for this effect by deriving a new angular separation distribution, $f(r)$, directly from the B(r) (Sec. \ref{sec:newfofr}). We show the new $f(r)$ in Figure \ref{fig:angdistVIK}, where we appear to still find statistically-significant VIKING galaxies at scales greater than 10 arcseconds. This large angular separation suggests that the lensing search in \cite{GN2012}, which only probes out to $\theta$ $\sim$3.5\", would have missed almost half of the likely counterparts.

The Einstein ring radius distribution from \cite{Aris2018} drops to zero beyond 1.5 arcseconds, unlike what is seen for our sample of lensed sources. This is smaller than the average positional uncertainty seen for SDSS counterparts in \cite{Bourne2014}, and both \cite{GN2014} and \cite{GN2017} find the contribution of strong lensing to be noticable to $\sim$10 arcseconds. The cross-correlation analysis in \cite{GN2017} models the effects of strong and weak gravitational lensing. They find that for angular separations larger than $\sim$12 arcseconds, weak gravitational lensing becomes the dominant cause for the observed cross-correlations.
Heavier halo masses, such as the ones associated with galaxy clusters instead of individual galaxies, create larger angular offsets than galaxy-galaxy lenses. It could be that our sample contains more galaxy-cluster lensing cases, accounting for the larger angular offsets we see \citep{GN2017,Aris2018}. These lensing cases were not seen in \cite{Aris2018}, which could be because our sample includes fainter sources, or because of the limited (15 galaxies) statistics of the study by \cite{Aris2018}.

\subsection{Extending the lens-selection method on the complete H-ATLAS catalogue}
\label{sec:discExtended}
\begin{figure}
\centering
\includegraphics[width=\linewidth]{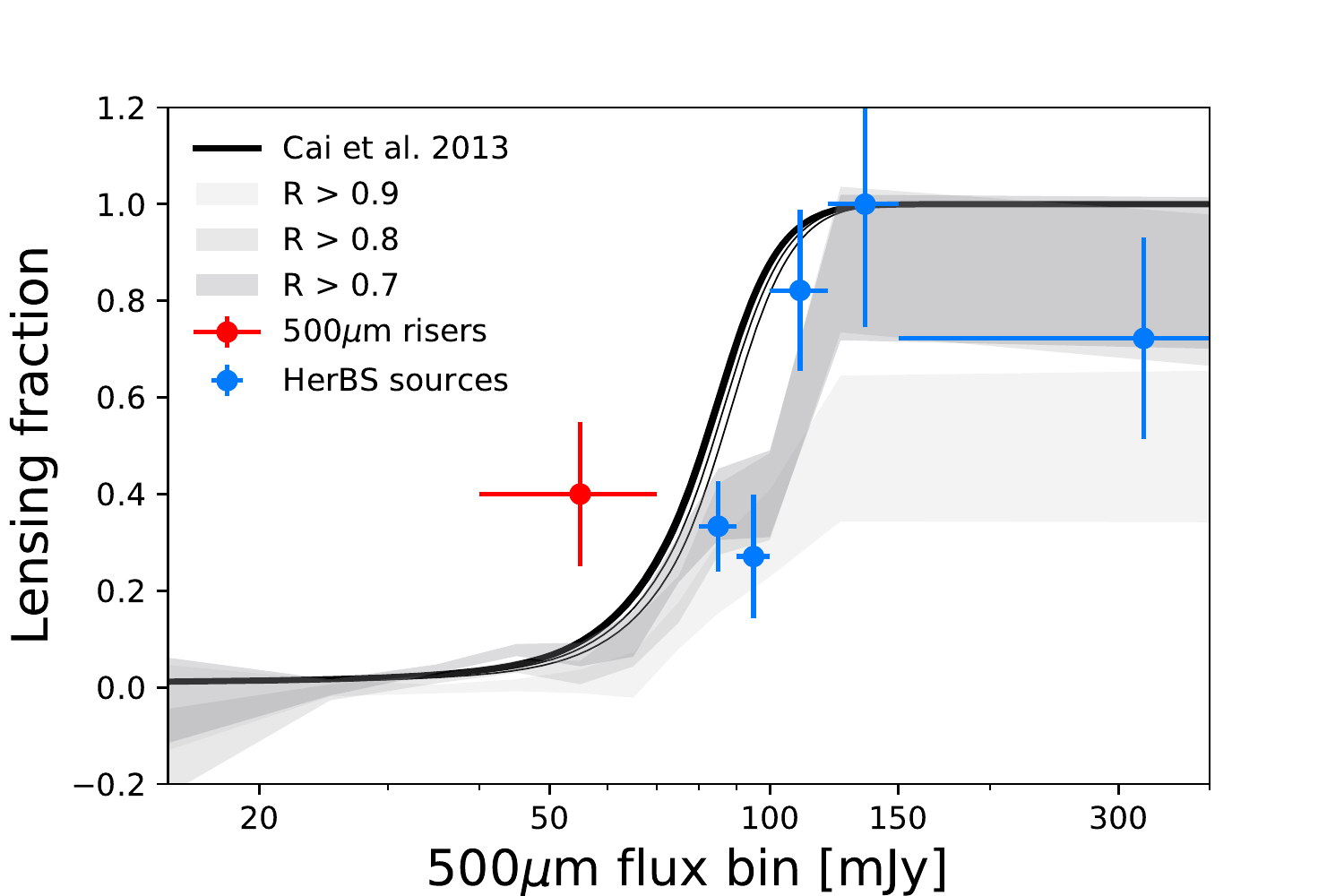}
\caption{The lensing fraction, corrected for false-positives, as a function of their 500$\mu$m selection flux \textcolor{referee}{density} shown for the entire H-ATLAS catalogue. The behaviour of the entire sample agrees with the HerBS sources. ALMA observations of the 500$\mu$m risers suggest 40\% are gravitationally-lensed, although our models suggest this is not true in general for \textit{Herschel} sources.}
\label{fig:lensFracGeneral}
\end{figure}
Finally, we test our adapted selection method on the complete H-ATLAS catalogue, by looking at the flux \textcolor{referee}{density}-evolution of the fraction of sources with reliable counterparts. Our initial HerBS sample was small, but the complete H-ATLAS catalogue consists of around half a million sub-mm sources. This large number of \textit{Herschel} sources allows us to make stringent cuts to the selection, and still achieve reasonable statistics.

We produce our sample in the following way. We start with the H-ATLAS catalogue, neglecting the NGP field, which does not have coverage in the VIKING fields. We remove sources with photometric redshifts smaller than 2 for the brightest \textit{Herschel} sources -- estimated by fitting the template from \citealt{Bakx2018}) -- and for sources fainter than 70 mJy at 500$\mu$m, we remove sources with photometric redshifts smaller than 3 to account for an increase in false-positives due to the increase in sample size. \textcolor{referee}{The template from \cite{Bakx2018} assumes a two-temperature modified blackbody SED. This template is fitted to the \textit{Herschel}/SPIRE (250, 350 and 500$\mu$m) and, where available, 850$\mu$m SCUBA-2 photometry of 24 \textit{Herschel} sources with spectroscopic redshifts greater than 1.5 and 500$\mu$m flux densities greater than 80mJy.} 
These redshifts are found by fitting the photometric template from \cite{Bakx2018}, similar to Section \ref{sec:vik_z_phot}. We remove sources within 10 arcseconds of an NVSS source, which removes all bright blazar objects, and potentially some of the brightest non-blazar sub-mm sources. This survey is also able to find all blazar-objects in the early HerBS catalogue \citep{Bakx2018}, and identified only three non-blazar sources. 

In Figure \ref{fig:lensFracGeneral} we show the lensing fraction, using equation \ref{eq:RatioToLF} with a counterparts selected at R > 0.7, 0.8 and 0.9 for each observed 500 $\mu$m flux \textcolor{referee}{density} bin, and also show the corrected VIKING analysis from Figure \ref{fig:lensFrac}. We also show the lensing fraction of the 500$\mu$m risers from \cite{Ivison2016} \citep{Oteo2017}. \textcolor{referee}{All flux densities are not corrected for magnification.}

We find good agreement between the models, HerBS sources and the full H-ATLAS catalogue. Sources with 500$\mu$m flux \textcolor{referee}{densities} around $\sim$90mJy also appear to result in lower lensing fractions, similar to seen for the HerBS sources. This is not entirely unexpected, since most sources at 500$\mu$m flux \textcolor{referee}{densities} are also included in the HerBS catalogue. We fail to explain why \cite{Oteo2017} found 40\% of their sources to be gravitationally-lensed. This could be due to their selection at high redshift (z$_{\rm mean}$ = 3.8), their non-trivial selection of sources by 'eyeballing', or because they identified their sources at 500$\mu$m, instead of the H-ATLAS catalogue \citep{Valiante2016}, which detected and extracted sources at their 250$\mu$m position.

\section{Conclusion}
\label{sec:conclusion}
We have
searched for foreground gravitational lenses for the sources
in the \textit{Herschel} Bright Source Survey (HerBS), a sample of
the brightest high-redshift \textit{Herschel} sources. A model predicts
that $\sim$76\% of the HerBS sources are subject to strong gravitational lensing \citep{Cai2013}, and observations suggest a significant portion of the brightest sources are gravitationally-lensed \citep{Negrello2016}. 

Using existing catalogues of counterparts to \textit{Herschel} sources found on the SDSS \textcolor{referee}{from \cite{Bourne2016}}, we found 31 probable lenses for 121 HerBS sources, much less than the prediction of the model. Even when we correct for lenses that are likely to be too far from the \textit{Herschel} sources to be found by the counterpart-identification procedure, the number of SDSS counterparts only increased to 41 out of 121 HerBS sources. This shows that the SDSS is not
deep enough to find all the lenses for high-redshift \textit{Herschel}
sources.

We carried out
our own search for lenses on the VIKING near-infrared survey, adapting the standard statistical method of finding counterparts to \textit{Herschel} sources to allow for the fact that most of the sources are probably lensed. We found probable lenses for 56 out of 98 HerBS sources. We were also able to show that, within 10 arcseconds, 82\% of the HerBS sources have foreground VIKING galaxies that associated with them, even if we were not able to identify the galaxy in all cases. We found that the overall fraction of lensed sources agrees well with the model. We also found tentative evidence for a decline in the fraction of lensed sources with decreasing 500-micron flux density, which is also a prediction of the model.

Using the VIKING data, we determined the distribution of distances between the \textit{Herschel} positions and the associated VIKING galaxy. We found that this distribution cannot be explained by astrometric errors, nor with a combination of astrometric errors and galaxy-galaxy lensing. This could indicate a larger contribution of galaxy-cluster lensing for fainter selection flux \textcolor{referee}{densities, as has been seen statistically by cross-correlation analysis \citep{GN2014,GN2017}.}

\section*{Acknowledgements}
TB and SAE have received funding from the European Union Seventh Framework Programme ([FP7/2007-2013] [FP\&/2007-2011]) under grant agreement No. 607254. We thank M. Negrello for his help in adjusting the lensing profiles of \cite{Cai2013} to $\mu_{\rm max}$ of 10, 20 and 30. We would like to thank the anonymous referee for their suggestions for both the structure and the contents of this paper.
The \textit{Herschel}-ATLAS is a project with \textit{Herschel}, which is an ESA space observatory with science instruments provided by European-led Principal Investigator consortia and with important participation from NASA. The \textit{Herschel}-ATLAS website is \url{http://www.h-atlas.org}. This work was supported by NAOJ ALMA Scientific Research Grant Number 2018-09B. 
Funding for the Sloan Digital Sky Survey IV has been provided by the Alfred P. Sloan Foundation, the U.S. Department of Energy Office of Science, and the Participating Institutions. SDSS-IV acknowledges
support and resources from the Center for High-Performance Computing at
the University of Utah. The SDSS web site is www.sdss.org.

SDSS-IV is managed by the Astrophysical Research Consortium for the 
Participating Institutions of the SDSS Collaboration including the 
Brazilian Participation Group, the Carnegie Institution for Science, 
Carnegie Mellon University, the Chilean Participation Group, the French Participation Group, Harvard-Smithsonian Center for Astrophysics, 
Instituto de Astrof\'isica de Canarias, The Johns Hopkins University, Kavli Institute for the Physics and Mathematics of the Universe (IPMU) / 
University of Tokyo, the Korean Participation Group, Lawrence Berkeley National Laboratory, 
Leibniz Institut f\"ur Astrophysik Potsdam (AIP),  
Max-Planck-Institut f\"ur Astronomie (MPIA Heidelberg), 
Max-Planck-Institut f\"ur Astrophysik (MPA Garching), 
Max-Planck-Institut f\"ur Extraterrestrische Physik (MPE), 
National Astronomical Observatories of China, New Mexico State University, 
New York University, University of Notre Dame, 
Observat\'ario Nacional / MCTI, The Ohio State University, 
Pennsylvania State University, Shanghai Astronomical Observatory, 
United Kingdom Participation Group,
Universidad Nacional Aut\'onoma de M\'exico, University of Arizona, 
University of Colorado Boulder, University of Oxford, University of Portsmouth, 
University of Utah, University of Virginia, University of Washington, University of Wisconsin, 
Vanderbilt University, and Yale University.



\bibliographystyle{mnras}
\bibliography{References} 


\appendix

\section{Table of VIKING counterparts}
This table contains the results of our VIKING cross-analysis. The HerBS number and sub-mm redshift are from \cite{Bakx2018}. R is the reliability, K$_{\rm{mag}}$ is the SExtractor-derived AUTO-magnitude and error. $\theta$ is the angular separation between the \textit{Herschel} and VIKING position, and the VIKING positions are of the most likely counterpart. z$_{\rm{VIK}}$ is either from \cite{Wright2018}, or found by using the EAZY-package (indicated with $\dagger$).
\begin{onecolumn}
\tablefirsthead{\toprule HerBS- & z$_{\rm{sub-mm}}$ & R & K$_{\rm{mag}}$ & $\theta$ [as] & RA$_{\rm{VIK}}$ & DEC$_{\rm{VIK}}$ & z$_{\rm{VIK}}$   \\ \midrule}
\tablehead{%
\multicolumn{8}{c}%
{{\bfseries  Continued from previous column}} \\
\toprule
HerBS- & z$_{\rm{sub-mm}}$ & R & K$_{\rm{mag, AB}}$ & $\theta$ [as] & RA$_{\rm{VIK}}$ [deg] & DEC$_{\rm{VIK}}$ [deg] & z$_{\rm{VIK}}$ \\ \midrule}
\label{tab:totalSources}
\tabletail{%
\midrule \multicolumn{8}{r}{{Continued on next page}} \\ \midrule}
\tablelasttail{%
\\\midrule}
\begin{center}
\begin{supertabular}{llllllll}
2 & 2.41 & 0.71  & 18.70 $\pm$ 0.04 & 0.35   & 176.65815 & -0.19216 & 0.658 $_{ 0.413} ^{ 0.727}$ \\
8 & 2.11 & 0.65  & 19.70 $\pm$ 0.07 & 2.82   & 132.38992 &  2.24578 & 0.649 $_{ 0.390} ^{ 1.201}$ \\
10 & 2.09 & 0.99  & 21.80 $\pm$ 0.12 & 0.82  & 173.85945 & -1.76835 & 0.765 $_{ 0.370} ^{ 0.761}$ \\
13 & 3.68 & 0.98  & 19.30 $\pm$ 0.03 & 0.73  & 216.05822 &  2.38446 & 0.670 $_{ 0.570} ^{ 0.710}$ $\dagger$\\
15 & 2.16 & 0.94  & 20.30 $\pm$ 0.09 & 1.45  & 213.46698 & -0.00749 & 0.810 $_{ 0.730} ^{ 0.870}$ $\dagger$\\
16 & 3.56 & 1.00  & 15.50 $\pm$ 0.01 & 0.87   & 212.51941 &  2.05196 & 0.477 $_{ 0.266} ^{ 0.483}$ \\
17 & 2.76 & 0.84  & 19.90 $\pm$ 0.07 & 3.66  & 351.38147 & -30.37711 & 0.650 $_{ 0.371} ^{ 1.214}$ \\
18 & 2.22 & 0.99  & 19.60 $\pm$ 0.04 & 0.05  & 351.08255 & -32.65736 & 0.647 $_{ 0.428} ^{ 0.742}$ \\
19 & 3.39 & 0.89  & 16.90 $\pm$ 0.01 & 1.91   & 135.79821 &  0.65180 & 0.654 $_{ 0.420} ^{ 0.751}$ \\
26 & 2.34 & 0.84  & 18.50 $\pm$ 0.03 & 2.44  & 344.68667 & -29.85625 & 0.472 $_{ 0.353} ^{ 0.723}$ \\
28 & 4.24 & 0.87  & 19.50 $\pm$ 0.04 & 1.95  & 347.06546 & -34.63352 & 0.840 $_{ 0.770} ^{ 0.880}$ $\dagger$\\
32 & 2.35 & 0.91  & 19.90 $\pm$ 0.06 & 2.59  & 139.66990 &  2.51281 & 0.470 $_{ 0.379} ^{ 0.731}$ \\
33 & 2.76 & 0.92  & 17.20 $\pm$ 0.01 & 1.57  & 342.02271 & -33.97207 & 0.655 $_{ 0.420} ^{ 0.750}$ \\
37 & 2.31 & 0.98  & 20.00 $\pm$ 0.06 & 0.85     & 351.59608 & -34.44533 & 0.475 $_{ 0.393} ^{ 0.722}$ \\
38 & 2.98 & 0.85  & 20.20 $\pm$ 0.06 & 1.45  & 221.53706 &  2.32474 & 1.000 $_{ 0.910} ^{ 1.080}$ $\dagger$\\
39 & 2.66 & 0.90  & 20.30 $\pm$ 0.06 & 3.13  & 352.25344 & -32.29590 & 0.654 $_{ 0.425} ^{ 0.730}$ \\
46 & 1.97 & 0.83  & 21.70 $\pm$ 0.10 & 1.99   & 221.48364 & -0.81426 & 0.765 $_{ 0.378} ^{ 0.749}$ \\
47 & 2.33 & 0.98  & 19.90 $\pm$ 0.07 & 2.16  & 343.21137 & -31.61665 & 0.656 $_{ 0.258} ^{ 0.703}$ \\
48 & 1.99 & 0.99  & 20.00 $\pm$ 0.06 & 1.30     & 183.25637 & -0.82314 & 0.320 $_{ 0.290} ^{ 0.570}$ $\dagger$\\
49 & 3.37 & 0.93  & 18.70 $\pm$ 0.02 & 1.83  & 346.44223 & -33.17759 & 0.620 $_{ 0.590} ^{ 0.660}$ $\dagger$\\
50 & 2.66 & 0.99  & 20.60 $\pm$ 0.07 & 0.82  & 180.82973 & -1.21517 & 0.765 $_{ 0.368} ^{ 0.751}$ \\
51 & 1.87 & 0.39  & 19.30 $\pm$ 0.04 & 4.63  & 181.78763 & -1.78506 & 1.080 $_{ 1.010} ^{ 1.150}$ $\dagger$\\
53 & 1.71 & 0.99  & 18.10 $\pm$ 0.02 & 0.52  & 177.80115 & -1.44379 & 0.550 $_{ 0.510} ^{ 0.570}$ $\dagger$\\
59 & 2.48 & 0.94  & 17.40 $\pm$ 0.01 & 2.16  & 138.27117 & -0.89519 & 0.654 $_{ 0.424} ^{ 0.755}$ \\
61 & 3.16 & 0.98  & 19.90 $\pm$ 0.05 & 1.40  & 180.36500 & -1.67919 & 0.662 $_{ 0.442} ^{ 0.706}$ \\
62 & 2.21 & 0.99  & 20.80 $\pm$ 0.09 & 0.90  & 183.92843 & -0.87221 & 0.980 $_{ 0.890} ^{ 1.070}$ $\dagger$\\
66 & 2.05 & 0.85  & 17.70 $\pm$ 0.01 & 8.95  & 179.58198 & -1.63277 & 0.005 $_{ 0.002} ^{ 0.009}$ \\
67 & 3.16   \\  
68 & 2.21 & 0.98  & 18.40 $\pm$ 0.02 & 0.83  & 339.47444 & -30.97466 & 0.654 $_{ 0.399} ^{ 0.757}$ \\
71 & 2.30 & 0.85  & 20.80 $\pm$ 0.10 & 6.38  & 173.18104 & -0.85340 & 0.661 $_{ 0.415} ^{ 0.732}$ \\
72 & 2.86 & 0.97  & 19.10 $\pm$ 0.03 & 0.85  & 221.30077 & -0.25289 & 0.473 $_{ 0.258} ^{ 0.704}$ \\
74 & 2.07 & 0.91  & 19.00 $\pm$ 0.10 & 6.82    & 181.50473 &  0.58381 & 0.652 $_{ 0.426} ^{ 0.755}$ \\
78 & 2.72 &  \\ 
80 & 2.02 & 0.44  & 21.40 $\pm$ 0.13 & 9.70  & 345.01197 & -31.83534 & 0.651 $_{ 0.423} ^{ 0.706}$ \\
82 & 2.23 & 1.00  & 18.80 $\pm$ 0.03 & 0.24  & 182.93699 &  1.11056 & 0.652 $_{ 0.370} ^{ 0.710}$ \\
83 & 3.93 & 0.57  & 20.20 $\pm$ 0.05 & 5.83  & 184.55209 &  1.31266 & 1.050 $_{ 1.000} ^{1.170} $ $\dagger$\\
84 & 2.39 & 0.63  & 20.00 $\pm$ 0.06 & 5.73     & 341.00463 & -34.00744 & 0.653 $_{ 0.426} ^{ 1.199}$ \\
85 & 2.15 & 0.86  & 19.80 $\pm$ 0.04 & 9.73  & 176.96765 & -0.97381 & 0.620 $_{ 0.560} ^{ 0.700}$ $\dagger$\\
91 & 1.79 & 0.72  & 19.50 $\pm$ 0.06 & 2.80  & 140.39949 &  0.02539 & 0.653 $_{ 0.249} ^{ 0.720}$ \\
96 & 2.09 & 0.93  & 19.00 $\pm$ 0.04 & 1.06   & 174.51500 & -1.29342 & 0.653 $_{ 0.395} ^{ 1.214}$ \\
97 & 2.28 & 0.58  & 19.40 $\pm$ 0.03 & 6.22  & 340.11721 & -34.52525 & 0.640 $_{ 0.590} ^{ 0.670}$ $\dagger$\\
99 & 2.26 & 0.69  & 19.80 $\pm$ 0.05 & 6.76  & 139.53995 &  0.32293 & 0.651 $_{ 0.426} ^{ 0.755}$ \\
100 & 1.94 & 0.61  & 20.10 $\pm$ 0.08 & 1.63 & 174.63876 &  0.81973 & 0.695 $_{ 0.405} ^{ 0.739}$ \\
103 & 2.15 & 0.73  & 21.00 $\pm$ 0.09 & 9.39  & 343.35065 & -32.58191 & 0.666 $_{ 0.394} ^{ 0.734}$ \\
105 & 2.48 & 0.60  & 19.10 $\pm$ 0.03 & 3.80 & 129.88328 & -1.30002 & 0.360 $_{ 0.320} ^{ 0.430}$ $\dagger$\\
108 & 2.76 & 0.99  & 19.70 $\pm$ 0.05 & 0.28 & 129.57267 & -0.69282 & 1.120 $_{ 1.050} ^{ 1.170}$ $\dagger$\\
110 & 2.84 & 0.52  & 18.10 $\pm$ 0.01 & 2.74  & 214.63674 &  1.03666 & 0.740 $_{ 0.70} ^{ 0.770}$ $\dagger$\\
111 & 2.10 & 0.99  & 20.10 $\pm$ 0.06 & 0.82 & 339.92667 & -33.55107 & 1.300 $_{ 1.210} ^{ 1.600}$ $\dagger$\\
116 & 3.20 & 0.83  & 20.70 $\pm$ 0.09 & 6.69 & 183.45014 &  1.13866 & 0.650 $_{ 0.244} ^{ 1.204}$ \\
119 & 2.38 & 0.42  & 19.90 $\pm$ 0.06 & 9.37 & 174.63934 & -1.78098 & 0.653 $_{ 0.442} ^{ 5.351}$ \\
121 & 3.11 &   \\
124 & 1.72 & 0.60  & 20.00 $\pm$ 0.10 & 7.30     & 185.49287 &  0.55890 & 0.440 $_{ 0.370} ^{ 0.490}$ $\dagger$ \\
126 & 2.44 &   \\
130 & 1.77 & 0.87  & 19.30 $\pm$ 0.07 & 7.06 & 216.77797 &  0.38112 & 1.090 $_{ 1.020} ^{ 1.160}$ $\dagger$\\
131 & 2.50 & 0.82  & 20.00 $\pm$ 0.06 & 5.82    & 343.41449 & -32.93117 & 0.478 $_{ 0.379} ^{ 1.059}$ \\
132 & 2.46 & 0.83  & 20.10 $\pm$ 0.06 & 2.12 & 348.02216 & -29.84062 & 0.652 $_{ 0.417} ^{ 0.73}$ \\
135 & 3.13 & 0.77  & 18.70 $\pm$ 0.04 & 7.56 & 344.05019 & -32.94930 & 0.640 $_{ 0.600} ^{ 0.660}$ $\dagger$\\
136 & 2.82 & 0.99  & 20.80 $\pm$ 0.07 & 0.61 & 133.28558 & -0.95793 & 0.473 $_{ 0.347} ^{ 0.732}$ \\
137 & 2.59 & 0.72  & 20.50 $\pm$ 0.06 & 6.17 & 223.40691 &  0.06889 & 1.046 $_{ 0.393} ^{ 1.242}$ \\
140 & 2.14 & 0.89  & 19.50 $\pm$ 0.04 & 3.34 & 215.41861 &  0.07911 & 0.675 $_{ 0.411} ^{ 0.732}$ \\
141 & 1.88 & 0.99  & 16.70 $\pm$ 0.01 & 1.76 & 341.99904 & -31.02638 & 0.653 $_{ 0.426} ^{ 1.041}$ \\
142 & 2.70 &   \\
143 & 2.14 & 0.92  & 21.20 $\pm$ 0.12 & 1.98 & 214.54221 & -0.62928 & 1.736 $_{ 0.516} ^{ 1.886}$ \\
146 & 2.01 & 1.00  & 18.50 $\pm$ 0.03 & 1.00 & 350.54548 & -33.63017 & 0.760 $_{ 0.720} ^{ 0.790}$ $\dagger$\\
147 & 2.22 & 0.40  & 20.30 $\pm$ 0.09 & 6.92 & 218.51328 &  0.04184 & 0.570 $_{ 0.530} ^{ 0.60}$ $\dagger$\\
148 & 2.01 & 0.93  & 21.50 $\pm$ 0.11 & 2.50 & 340.11011 & -31.86570 & 1.280 $_{ 0.710} ^{ 1.360}$ $\dagger$\\
150 & 3.36 & 0.69  & 21.10 $\pm$ 0.09 & 9.22 & 186.24910 & -0.94609 & 0.667 $_{ 0.411} ^{ 0.73}$ \\
153 & 2.02 & 0.84  & 19.00 $\pm$ 0.03 & 1.61  & 220.68120 &  1.91828 & 0.650 $_{ 0.600} ^{ 0.670}$ $\dagger$\\
157 & 2.66 & 0.43  & 20.90 $\pm$ 0.10 & 6.95 & 132.48967 &  1.12223 & 1.030 $_{ 0.990} ^{ 1.070}$ $\dagger$\\
161 & 2.22 &    \\
162 & 2.63 & 0.74  & 21.00 $\pm$ 0.11 & 3.64  & 220.89357 & -0.50866 & 0.930 $_{ 0.750} ^{ 1.070}$ $\dagger$\\
164 & 2.26 & 0.98  & 20.80 $\pm$ 0.10 & 1.67  & 183.56782 & -1.61769 & 0.646 $_{ 0.265} ^{ 1.044}$ \\
165 & 2.39 &   \\
168 & 3.87 & 1.00  & 18.20 $\pm$ 0.02 & 0.54 & 342.68950 & -30.78877 & 0.470 $_{ 0.430} ^{ 0.490}$ $\dagger$\\
169 & 2.40 &   \\
171 & 2.33 & 0.58  & 20.60 $\pm$ 0.11 & 1.85 & 129.93774 &  2.17226 & 0.768 $_{ 0.403} ^{ 0.750}$ \\
172 & 2.16 &  \\
175 & 2.80 &  \\
177 & 3.93 & 0.99  & 20.20 $\pm$ 0.10 & 0.52  & 178.64031 &  0.84501 & 0.690 $_{ 0.630} ^{ 0.730}$ $\dagger$\\
179 & 3.08 & 0.95  & 21.20 $\pm$ 0.13 & 0.51 & 178.83745 & -2.22485 & 0.653 $_{ 0.434} ^{ 0.732}$ \\
182 & 2.92 & 0.75  & 21.30 $\pm$ 0.11 & 4.18 & 346.41164 & -31.36818 & 0.778 $_{ 0.394} ^{ 1.039}$ \\
183 & 2.61 & 0.68  & 18.80 $\pm$ 0.03 & 1.62 & 136.22176 &  2.33790 & 0.656 $_{ 0.410} ^{ 0.715}$ \\
185 & 3.05 & 0.92  & 18.10 $\pm$ 0.01 & 2.12  & 141.03735 & -0.83826 & 0.651 $_{ 0.414} ^{ 1.191}$ \\
187 & 1.86 & 0.82   \\
188 & 2.03 & 0.65  & 20.80 $\pm$ 0.10 & 6.70  & 130.74862 &  2.83452 & 0.650 $_{ 0.429} ^{ 1.207}$ \\
189 & 2.18 & 0.95  & 19.00 $\pm$ 0.04 & 1.65  & 344.00331 & -31.54221 & 0.672 $_{ 0.406} ^{ 0.724}$ \\
190 & 2.54 & 0.51  & 19.90 $\pm$ 0.06 & 9.54 & 136.02100 & -0.55869 & 0.664 $_{ 0.414} ^{ 0.735}$ \\
193 & 2.54 & 0.85  & 19.00 $\pm$ 0.04 & 9.15  & 133.46455 & -0.13364 & 0.840 $_{ 0.800} ^{ 0.880}$ $\dagger$\\
194 & 2.26 & 0.93  & 17.40 $\pm$ 0.01 & 6.89  & 133.83948 & -0.59959 & 0.651 $_{ 0.425} ^{ 1.210}$ \\
195 & 2.35 & 0.64  & 17.60 $\pm$ 0.01 & 9.11 & 224.47376 &  0.00396 & 0.655 $_{ 0.428} ^{ 1.199}$ \\
197 & 2.04 &    \\
201 & 2.92 & 0.64  & 20.20 $\pm$ 0.08 & 8.70 & 212.82686 & -1.11476 & 0.474 $_{ 0.414} ^{ 0.737}$ \\
202 & 1.87 & 0.60  & 20.50 $\pm$ 0.12 & 8.91 & 218.37044 &  2.13505 & 0.652 $_{ 0.424} ^{ 0.712}$ \\
203 & 1.71 & 0.99  & 16.50 $\pm$ 0.01 & 2.01 & 214.61496 & -0.28455 & 0.654 $_{ 0.424} ^{ 0.756}$ \\
205 & 2.59 & 0.86  & 19.30 $\pm$ 0.03 & 4.69 & 222.88625 &  2.68495 & 0.651 $_{ 0.413} ^{ 0.703}$ \\
206 & 1.98 &    \\
208 & 3.59 &    \\
209 & 2.89 & 0.83  & 21.50 $\pm$ 0.15 & 9.87 & 342.33542 & -33.49470 & 0.508 $_{ 0.427} ^{ 0.741}$  $\dagger$
\end{supertabular}%
\end{center}
\end{onecolumn}


\bsp	
\label{lastpage}
\end{document}